\documentclass[a4paper,11pt]{article}
\pdfoutput=1 

\usepackage{jcappub} 

\usepackage[T1]{fontenc} 
\title{\boldmath Sensitivity of CTA to dark matter signals from the Galactic Center}

\author[a,b,1]{Mathias Pierre\note{Corresponding author.},}
\author[c,2]{Jennifer M.~Siegal-Gaskins\note{Einstein fellow.},}
\author[d,e]{and Pat Scott}

\affiliation[a]{\'{E}cole Normale Sup\'{e}rieure de Cachan,\\61 Avenue du Pr\'{e}sident Wilson, 94230 Cachan, France}
\affiliation[b]{Universit\'{e} Pierre et Marie Curie,\\4 Place Jussieu, 75005 Paris, France}
\affiliation[c]{California Institute of Technology,\\1200 E.~California Blvd., Pasadena, CA 91125, USA}
\affiliation[d]{Department of Physics, McGill University,\\ 3600 Rue University, Montr\'{e}al, Qu\'{e}bec, H3A 2T8, Canada}
\affiliation[e]{School of Physics, The University of Sydney,\\ NSW 2006, Australia}

\emailAdd{mathias.pierre@ens-cachan.fr}
\emailAdd{jsg@tapir.caltech.edu}
\emailAdd{patscott@physics.mcgill.ca}

\abstract{The Galactic Center is one of the most promising targets for indirect detection of dark matter with gamma rays.  We investigate the sensitivity of the upcoming Cherenkov Telescope Array (CTA) to dark matter annihilation and decay in the Galactic Center.  As the inner density profile of the Milky Way's dark matter halo is uncertain, we study the impact of the slope of the Galactic density profile, inwards of the Sun, on the prospects for detecting a dark matter signal with CTA\@.   Adopting the Ring Method to define the signal and background regions in an ON-OFF analysis approach, we find that the sensitivity achieved by CTA to annihilation signals is strongly dependent on the inner profile slope, whereas the dependence is more mild in the case of dark matter decay.  Surprisingly, we find that the optimal choice of signal and background regions is virtually independent of the assumed density profile.  For the fiducial case of a Navarro-Frenk-White profile, 
we find that CTA will be able to probe annihilation cross sections well below the canonical thermal relic value for dark matter masses from a few tens of GeV up to $\sim 5$~TeV for annihilation to $\tau^{+}\tau^{-}$, and will achieve only a slightly weaker sensitivity for annihilation to $b\bar{b}$ or $\mu^{+}\mu^{-}$.  CTA will improve significantly on current sensitivity to annihilation signals for dark matter masses above $\sim 100$~GeV, covering parameter space that is complementary to that probed by searches with the Fermi Large Area Telescope.  The interpretation of apparent excesses in the measured cosmic-ray electron and positron spectra as signals of dark matter decay will also be testable with CTA\@ observations of the Galactic Center.  We demonstrate that both for annihilation and for decay, including spectral information for hard channels (such as $\mu^{+}\mu^{-}$ and $\tau^{+}\tau^{-}$) leads to enhanced sensitivity for dark matter masses above $m_{\rm DM}\sim 200$~GeV.}

\begin{document}
\maketitle
\flushbottom

\section{Introduction}
\label{sec:intro}

The existence of dark matter (DM) on Galactic and cosmological scales is well established \cite{Bertone:2004pz,Ade:2013zuv}, however its detailed particle properties are still unknown.  Many candidate DM  particles can annihilate or decay to Standard Model particles, and thus may be detected indirectly through searches for the annihilation or decay products.  Indirect searches in gamma rays \cite{BringmannWeniger} are well suited to detect weakly interacting massive particles (WIMPs), which have masses in the GeV to multi-TeV range and produce photon signals of similar energies.  If DM consists entirely of WIMPs produced thermally in the early universe, in the simplest case this implies that the thermally-averaged product of their pair-annihilation cross section and relative velocity $\langle \sigma v \rangle $ is a few times $10^{-26}$~cm$^{3}$~s$^{-1}$~\cite{Jungman:1995df,Steigman:2012nb} (although in some scenarios thermal production is viable with $\langle \sigma v \rangle $ up to two orders of magnitude larger or smaller than this).  Indirect searches in gamma rays by the Fermi Large Area Telescope (LAT)~(e.g.~\cite{Ackermann:2013yva,Hooper:2012sr}) have recently reached the sensitivity required to test this canonical annihilation cross section for low WIMP masses.

The Galactic Center (GC) has long been identified as a promising target for indirect DM searches, due to its close proximity and high concentration of DM~(e.g.~\cite{Bergstrom:1997fj}).  Searches for DM signals from the GC in gamma rays have generated some of the strongest constraints to date on DM models~\cite{Hooper:2012sr,Gomez-Vargas:2013bea}, however a major uncertainty in interpreting the results of indirect searches in the GC is the DM distribution. 

DM-only N-body simulations of structure formation have long suggested a universal density profile for DM halos spanning a large range of masses.  For many years the two-parameter Navarro-Frenk-White (NFW) profile~\cite{Navarro:1995iw} was the canonical choice, however, results of recent higher-resolution simulations have favored the Einasto density profile~\cite{1965TrAlm...5...87E}, which has been found to provide an improved fit to DM halo profiles at the expense of adding an additional parameter~\cite{nfwsmooth,Navarro:2008kc}.

Observational constraints on the DM distribution near the GC are weak, as the gravitational potential is dominated by baryons in the Inner Galaxy.  Arguing against the steeper NFW and Einasto profiles, some evidence from the rotation curves of spiral galaxies favors cored DM profiles~(e.g.~\cite{Persic:1995ru, Gentile:2004tb}).  In addition, it has been pointed out that the presence of baryons may significantly alter the DM distribution through various mechanisms.  These include adiabatic contraction~(e.g. \cite{gondolosilk,Gnedin:2004cx,Gnedin:2011uj}), which steepens the DM distribution near the GC, enhancing gamma-ray and other indirect signals from the GC~\cite{gondolosilk,Prada:2004pi,Gustafsson:2006gr,Gomez-Vargas:2013bea,Lacroix:2013qka}, and feedback~\cite{Mashchenko:2007jp, Maccio:2011eh,Pontzen:2011ty, Governato:2012fa}, which tends to make the DM density profile flatter.  Modifying DM particle properties, e.g.~via adding self-interactions, can also flatten the DM density profile~\cite{Rocha:2012jg,  Kaplinghat:2013xca}.  Varying the inner slope of the DM density profile has a substantial impact on the predicted indirect signals.

Although the Fermi LAT has enabled very sensitive searches for low-mass WIMPs, its sensitivity is weaker for higher-mass WIMPs.  The LAT detects gamma rays from $\sim 20$~MeV to $> 300$~GeV, becoming statistics-limited at high energy.  Complementary to the LAT in energy, ground-based imaging atmospheric Cherenkov telescopes (IACTs) are sensitive to gamma rays with energies above $\sim 100$~GeV, and feature much larger effective areas than the LAT\@.  The current generation of IACTs, including H.E.S.S., MAGIC, and VERITAS, have performed DM searches, but, apart from WIMPs that produce substantial virtual internal bremsstrahlung \cite{Ripken11}, have been so far unable to test favored regions of parameter space for DM annihilation.  The Cherenkov Telescope Array (CTA)~\cite{Acharya:2013sxa}, a future IACT, will provide a substantial improvement in DM indirect detection capabilities \cite{Bergstrom11,Doro:2012xx,Wood:2013taa,Cline13b}.  This is in large part due to its larger effective area and field of view in comparison to current IACTs.  CTA will also be sensitive to gamma rays with energies as low as $\sim 20$~GeV, extending the energy threshold far below that of current IACTs and providing significant overlap with the Fermi LAT energy range.
 
With its enhanced sensitivity at high energies, CTA will also have the potential to test DM interpretations of apparent anomalies in the measured local cosmic-ray fluxes by searching for associated gamma-ray signals.  The recent precise measurement of the local positron fraction by AMS-02~\cite{Aguilar:2013qda} again confirmed the rise in the positron fraction above a few GeV which had been previously observed by PAMELA~\cite{Adriani:2008zr} and subsequently by the Fermi LAT~\cite{FermiLAT:2011ab}.  With the AMS-02 data, the rise in the positron fraction is now observed to extend to $\sim 300$~GeV, and no turnover has been identified.  The total cosmic ray electron (CRE; used here to refer to electron plus positron) spectrum has also been measured by several experiments (e.g.~\cite{Chang:2008aa,Abdo:2009zk,Aharonian:2008aa,Aharonian:2009ah}).  These measurements indicate that the total CRE spectrum scales roughly as $E^{-3}$ above a few tens of GeV, and reveal a slight excess over conventional expectations at a few hundred GeV, with the spectrum turning over and falling above $\sim 1$~TeV.  Together with the positron fraction measurement, this points to a DM mass of at least several hundred GeV if DM annihilation or decay is invoked as the origin of the excesses (e.g.~\cite{Cirelli:2012ut}); this mass range is easily accessible to CTA\@.  Constraints on these scenarios have been obtained from a variety of gamma-ray measurements, including H.E.S.S. observations of the GC~\cite{Abazajian:2011ak} and Fornax~\cite{Cirelli:2012ut} and Fermi LAT observations of the isotropic gamma-ray background (IGRB)~\cite{Cirelli:2012ut,Abdo:2010dk} and the Milky Way halo~\cite{Ackermann:2012rg}.  Although scenarios producing the cosmic-ray excesses by DM annihilation are now largely excluded, some parameter space remains viable for interpretations in terms of DM decay.

In this work we investigate the sensitivity of CTA to gamma-ray signals from DM annihilation and decay in the GC\@.  Whereas previous studies have examined CTA's sensitivity to gamma rays from annihilation for benchmark density profiles, here we consider both annihilation and decay signals, as well as variations in the DM density profile.  The impact of variations in the density profile for GC searches with CTA is somewhat complex, as the sensitivity is not simply tied to the amplitude of the flux from a chosen signal region.  This is due to the need for IACTs, such as CTA, to use a background control region (``OFF'' region) to directly measure the intensity of the cosmic-ray background.  For a GC observation, in practice there will be some DM signal in the OFF region.  For this reason, CTA is sensitive only to \emph{variations} in the intensity of the signal between the chosen signal and background regions, and thus the sensitivity depends on both the slope of the DM density profile and the overall amplitude of the signal.  We evaluate the impact of varying the DM distribution for a wide range of possible density profiles, and study the issue of optimizing the ON and OFF regions for different assumed profiles.   Prior work focused on CTA has most often considered searches for DM-induced excesses integrated over broad energy ranges (see~\cite{Bringmann:2011ye} for an exception).  Here we also study the impact of spectral analysis, which exploits the different spectral shapes of the signal and background.  Spectral analysis was recently shown to improve sensitivity in a DM search using MAGIC~\cite{Aleksic:2013xea}.

This paper is structured as follows.  In \S\ref{sec:signals} we present the calculation of the gamma-ray signals from annihilation and decay, and define the DM density profiles used in this work.  The observational setup assumed for CTA is described in \S\ref{sec:obs}, and the likelihood analysis in \S\ref{sec:likelihood}.  Our results are given in \S\ref{sec:results}.  We discuss and conclude in \S\ref{sec:conc}.

\section{Dark matter signals from the Galactic Center}
\label{sec:signals}

\subsection{Gamma-ray intensity}
The differential gamma-ray intensity (photons per area per time per solid angle per energy) from annihilation of two DM particles $\chi$ in the GC is given by
\begin{equation}
\frac{d\Phi_{\rm ann}}{d\Omega\,dE} =\frac{\langle \sigma v \rangle}{8\pi m_{\chi}^{2}} \frac{dN_{\gamma}}{dE} \underbrace{\int_{\rm los}{\rho_{\chi}^{2}(r) dl}}_{J_{\rm ann}}
\end{equation}
where $\langle \sigma v \rangle$ is the average annihilation cross section times relative velocity, $m_{\chi}$ is the mass of the DM particle, and $dN_{\gamma}/dE$ is the energy spectrum of photons emitted per annihilation.  The function $\rho_{\chi}(r)$ is the DM density as a function of the distance $r$ from the GC\@. The coordinate $l$ runs along the line-of-sight (los), and $r(l,\psi)=\sqrt{r_{\odot}^{2}+l^{2}-2r_{\odot}l\cos(\psi)}$ where $r_{\odot}$ is the distance between the Sun and the GC, and $\psi$ the angle between the line-of-sight and the direction of the GC\@.  The line-of-sight integral of the DM density squared is often referred to as the ``astrophysical factor'' or ``J factor'' and is denoted $J_{\rm ann}$, defined here as differential in solid angle.

For the case of DM decay, the differential gamma-ray intensity is given by
\begin{equation}
\frac{d\Phi_{\rm dec}}{d\Omega\,dE} =\frac{1}{4\pi \tau m_{\chi}} \frac{dN_{\gamma}}{dE} \underbrace{\int_{\rm los}{\rho_{\chi}(r) dl}}_{J_{\rm dec}}
\end{equation}
where $\tau$ is the lifetime of the DM particles, and here $dN_{\gamma}/dE$ is the energy spectrum of photons emitted per decay.  The ``astrophysical factor'' for decay $J_{\rm dec}$ is given by the line-of-sight integral over the DM density.

The energy spectrum of the photons produced by DM annihilation or decay can be written as a sum over all possible final states
\begin{equation}
\frac{dN_{\gamma}}{dE}=\sum_{f} B_{f}\frac{dN_{f}}{dE}
\end{equation}
where $B_{f}$ is the branching fraction of final state $f$, and $dN_{f}/dE$ is the photon spectrum from annihilation or decay to the final state $f$. We calculate the photon spectrum for each final state using \textsc{DarkSUSY}~\cite{Gondolo:2004sc}.

\subsection{Density profiles}

In light of the uncertainty in the DM distribution in the Inner Galaxy and the dependence of the signal on it, we consider a range of density profiles.
As benchmarks, we consider the widely-used NFW profile~\cite{Navarro:1995iw}, as well as the Einasto profile~\cite{1965TrAlm...5...87E}.  The NFW profile is given by
\begin{equation}
\rho_{\rm NFW}(r)\propto \frac{1}{\left(\frac{r}{r_{s}}\right)\left[1+\left(\frac{r}{r_{s}}\right)\right]^{2}},
\end{equation}
where $r$ is the distance from the GC and $r_{s}$ is a scale radius, which indicates the transition radius between where the density scales as $r^{-1}$ and $r^{-3}$.  We fix $r_{s}=20$~kpc~\cite{Navarro:2008kc}.  
The Einasto profile is described by
\begin{equation}
\rho_{\rm Ein}(r)\propto e^{-\left(\frac{2}{a}\right) \left[ \left( \frac{r}{r_{s}} \right)^{a}-1 \right] },
\end{equation}
where $r_{s}$ is a scale radius and $a$ controls how rapidly the profile slope changes.
We take $r_{s}=20$~kpc and $a=0.17$~\cite{Navarro:2008kc}.

To consider flatter and steeper profiles, we generalize the NFW profile as follows
\begin{equation}
\rho_{\gamma}(r)\propto \frac{1}{\left(\frac{r}{r_{s}}\right)^{\gamma}\left[1+\left(\frac{r}{r_{s}}\right)\right]^{3-\gamma}}
\end{equation}
and vary the inner slope $\gamma$ from 0.5 to 1.5, fixing $r_{s}=20$~kpc as for the NFW and Einasto profiles.  The NFW profile corresponds to $\gamma=1$.
For all density profiles, we normalize the density at the solar circle to $\rho_{\odot}=0.43$ GeV/cm$^{3}$~\cite{Salucci:2010qr}.

\begin{figure}[h!]
	\begin{center}
	\includegraphics[width=8cm]{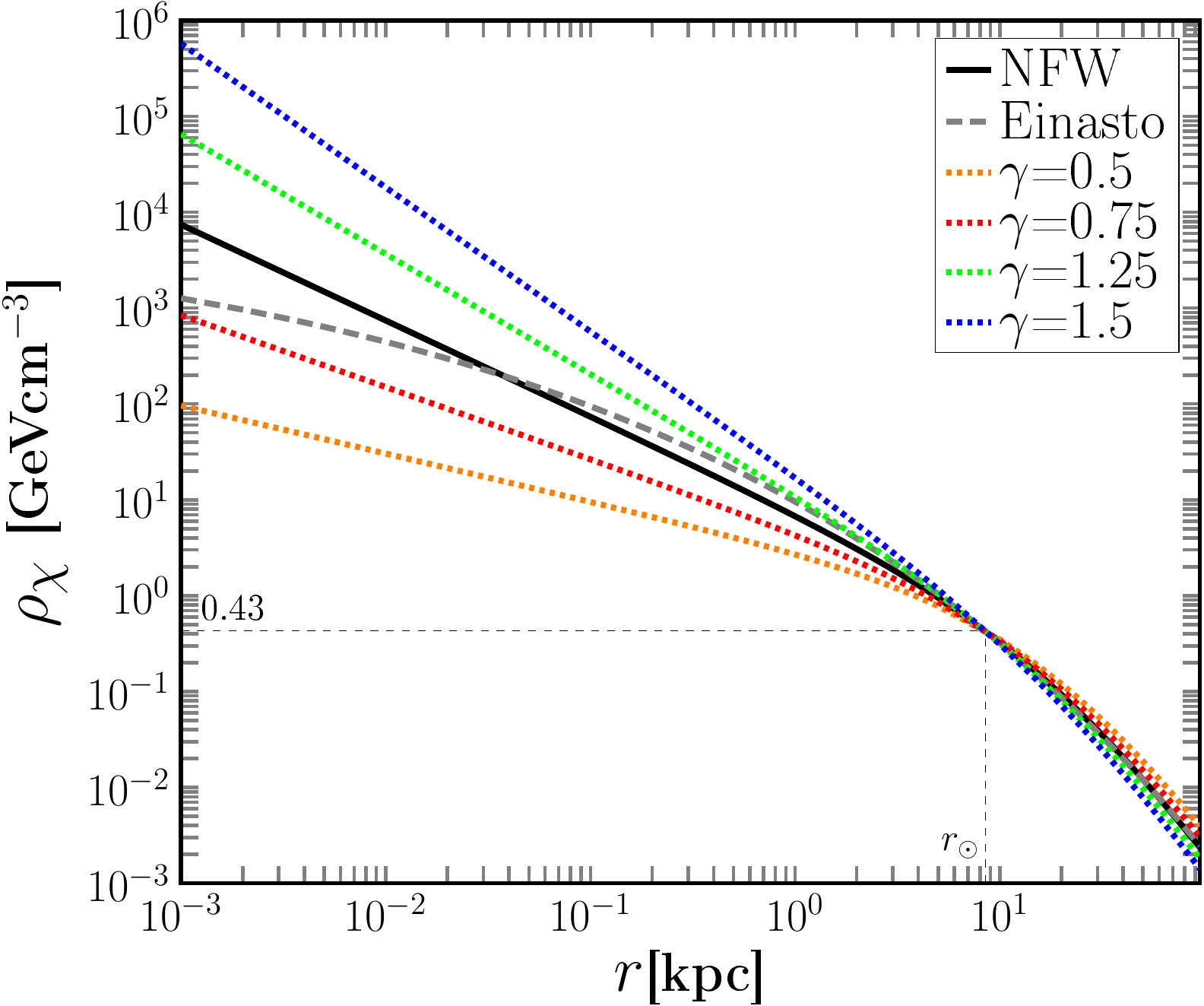}
	\caption{Variation of the DM density with the distance from the GC for different density profiles. 
	\label{fig:profiles}}
	\end{center}
\end{figure}

\begin{figure}[h!]
\centering
	\includegraphics[width=7cm]{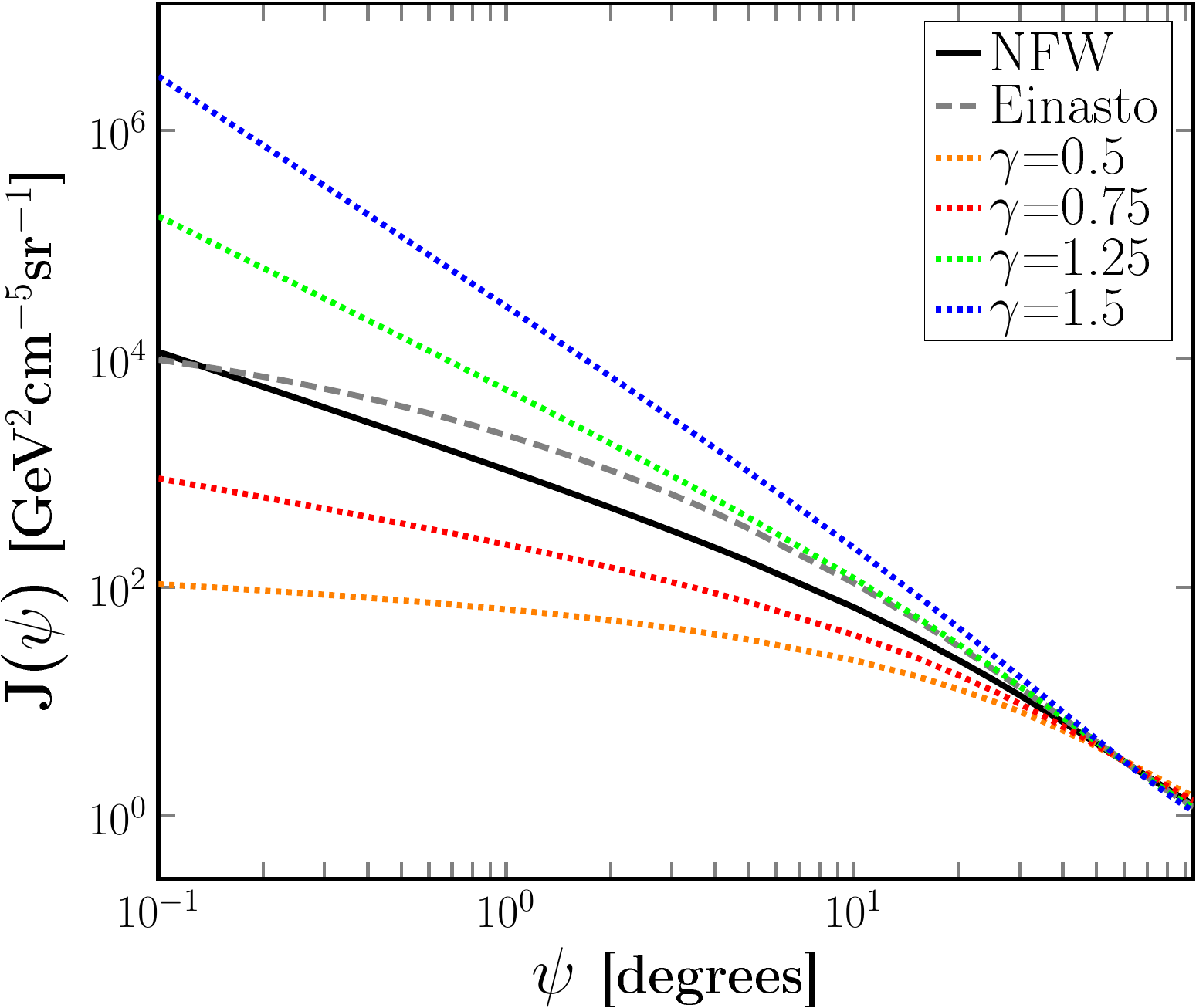}\hspace{.5cm}
	\includegraphics[width=7cm]{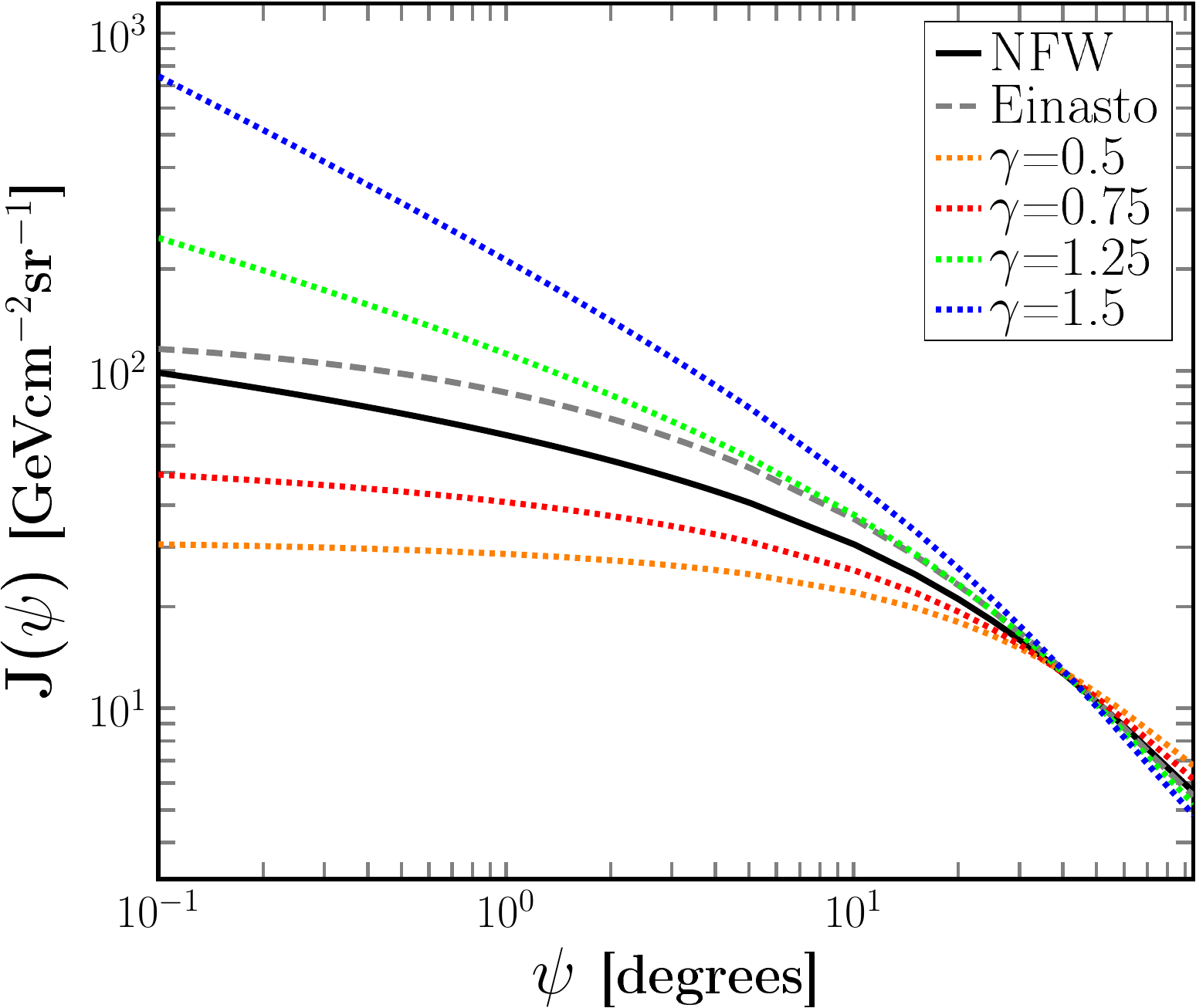}
\caption{Variation of the $J$ factor as a function of the angle between the line-of-sight and the GC for annihilation (left) and decay (right).
\label{fig:jfactors}}
\end{figure}

The density profiles considered here are compared in Fig.~\ref{fig:profiles}.  Fig.~\ref{fig:jfactors} shows the corresponding $J$ factors for annihilation and decay as a function of angle from the GC\@.  Near the GC the variation between $J$ factors for different profiles for annihilation spans many more orders of magnitude than the variation in the case of decay.

\section{Observational setup}
\label{sec:obs}

\subsection{Observational capabilities of CTA}

The CTA collaboration has explored the observational capabilities of several different possible array configurations~\cite{Bernlohr:2012we}.  In this work, we adopt the simulated results for Array I, which consists of 3 large size telescopes, 18 medium size telescopes and 56 small size telescopes.   We use the effective area and energy resolution obtained by the Paris-MVA analysis method assuming Array I~\cite{Bernlohr:2012we}.  Several other array configurations have been adopted in other analyses, in particular Arrays E and B were considered in the GC sensitivity analysis of~\cite{Doro:2012xx}.   Arrays I and E are both balanced in terms of having a range of telescope sizes and good sensitivity over a broad energy range.  Array B is an example of a compact array, which is more optimized for low energies and for DM studies tends to yield better sensitivity (as in~\cite{Doro:2012xx}).  In this work we adopt Array I due to the fact that detailed information about its expected capabilities that is necessary for this study (i.e., effective area and energy resolution) is readily available.  We thus expect that our analysis will yield conservative results with respect to studies that adopt Array B.

CREs represent an irreducible background for gamma-ray observations with IACTs such as CTA, as CRE-induced atmospheric showers cannot be distinguished from gamma-induced showers.
For this study, we consider CREs as an isotropic background.  The CRE spectrum has been measured by the Fermi LAT from $\sim20$~GeV up to $\sim1$~TeV~\cite{Abdo:2009zk}, and is well approximated by a power law $\propto E^{-3}$, which we use to describe the CRE background at all energies considered.  Specifically we take $E^{3}\, dN_{\rm CRE}/dE\,dA\,dt\,d\Omega=1.5 \times 10^{-2}$~GeV$^{2}$~cm$^{-2}$~s$^{-1}$~sr$^{-1}$.

In general IACTs can reject hadronic showers with high efficiency, thus we treat this possible source of background as negligible.  This is a somewhat optimistic treatment, as simulations find that the hadronic background for CTA is typically comparable to the CRE background at most energies and may dominate at low energies ($\lesssim 200$~GeV) and very high energies ($\gtrsim$~several TeV)~\cite{Bernlohr:2012we}.  However, a straightforward estimate of its magnitude for an analysis such as the one considered in this study is not currently available.  This issue warrants further dedicated study, but is expected to degrade limits by less than a factor of a few at all masses considered here.

\subsection{Signal and background regions}
\label{sec:ring}

\begin{figure}[ht]
\centering
	\includegraphics[width=6cm]{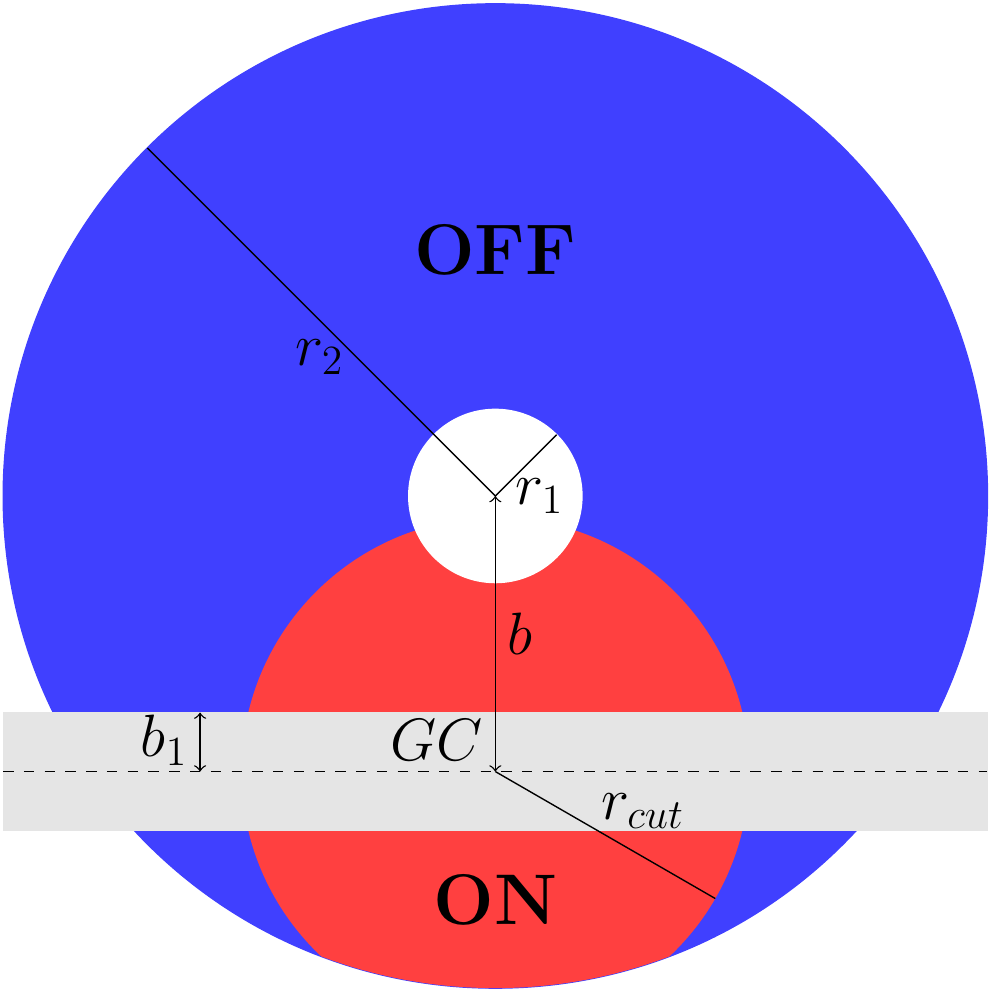}
\caption{Illustration of the choice of ON and OFF regions.  The ON and OFF regions are chosen within a ring centered on the FOV with inner radius $r_1$ and outer radius $r_2$.  For the GC observation considered here, the center of the FOV is offset by $b$ degrees in latitude from the GC\@.  The ON region is shown in red, defined by the intersection of a circle of radius $r_{\rm cut}$ centered on the GC and the ring with inner radius $r_1$ and outer radius $r_2$.  The OFF region, shown in blue, is defined by the remainder of the ring outside of the ON region.  The Galactic plane is excluded by a latitude cut of $b_1$ degrees (shown by the gray rectangle) from both the ON and OFF regions.  
\label{fig:ring}}
\end{figure}

To search for a DM signal we define a signal region (denoted ON) and background region (denoted OFF) within the field of view (FOV) of CTA using the \emph{Ring Method}~\cite{Doro:2012xx}.  This method is used to search for an excess over an isotropic background such as CRE-induced showers or hadronic showers.  The ON and OFF regions are illustrated in Fig.~\ref{fig:ring}, and are chosen to lie within a ring centered on the FOV of CTA\@.   This geometry is chosen to reduce systematics associated with variation of the effective area across the FOV\@.  For our default analysis we adopt the optimized parameters for the Array B reported in~\cite{Doro:2012xx} for an NFW profile: the inner and outer radii of the ring $r_1=0.44^\circ$ and $r_2=2.50^\circ$, the offset of the center of the ON region from the GC $b=1.4^\circ$, and the radius of the ON region $r_{\rm cut}=1.29^\circ$; we explore the dependence of the sensitivity on the ON and OFF region parameters for variations in the inner slope of the DM density profile in \S\ref{sec:results}.  We adopt the same ON and OFF regions for all DM masses.  This choice of ON and OFF regions requires a FOV of only $5^\circ$, however we note that the FOV of CTA increases at high energies to up to $\sim~10^\circ$, and thus this analysis could in principle be optimized for large DM masses to take advantage of the increased FOV\@. 

The Galactic plane is excluded within $|b_1|<0.3^\circ$ to avoid non-DM astrophysical gamma-ray emission; observations by current IACTs have shown that there is negligible diffuse astrophysical signal outside of $|b|<0.3$ degrees~\cite{Doro:2012xx}, although at the lowest energies considered here this may be a slightly optimistic treatment.  Point sources identified in the ON or OFF regions are assumed to be masked.  Note that a DM signal is present in both the ON and OFF regions~\cite{Mack:2008wu}, but will be larger (per unit solid angle) in the ON region.  

For the parameters adopted here, the angular size of the ON region is $\Delta \Omega_{\rm ON} = 9.9 \times 10^{-4}$~sr, while the angular size of the OFF region is $\Delta \Omega_{\rm OFF} = 4.0 \times 10^{-3}$~sr. We define the geometrical parameter $\alpha=\Delta \Omega_{\rm ON}/\Delta \Omega_{\rm OFF}$, which is the ratio of the solid angles of the ON and OFF regions.

The number of photons observed from a specified region of the sky from DM annihilation is
\begin{equation}
\label{eq:nann}
N_{\rm ann}=t_{\rm obs}\frac{\langle \sigma v \rangle}{8\pi m_{\chi}^{2}} N_{\gamma,{\rm obs}} \int_{\Delta \Omega}J_{\rm ann}(\psi)d\Omega,
\end{equation}
while for the case of DM decay, the number of observed photons is 
\begin{equation}
N_{\rm dec}=t_{\rm obs}\frac{1}{4\pi \tau m_{\chi}} N_{\gamma,{\rm obs}} \int_{\Delta \Omega}J_{\rm dec}(\psi)d\Omega,
\end{equation}
where $t_{\rm obs}$ is the observation time, and
\begin{equation}  
N_{\gamma,{\rm obs}}=
\int_{\Delta E}
{\int_{-\infty}^{+\infty}\frac{dN_{\gamma}(\bar{E})}{dE}A_{\rm eff}(\bar{E})\frac{e^{-\frac{(E-\bar{E})^{2}}{2\sigma^{2}}}}{\sqrt{2\pi\sigma^{2}}}d\bar{E}
dE},
\end{equation}
where $A_{\rm eff}(E)$ is the energy-dependent effective area.  The quantity $N_{\gamma,{\rm obs}}$ is the energy spectrum per annihilation or decay multiplied by the effective area of CTA and convolved with its energy resolution, integrated over the energy range considered ($\Delta E$).  Here we have modeled the energy resolution of CTA by convolving the source energy spectra with a Gaussian with energy-dependent width $\sigma(E)$, again assuming the capabilities obtained by the Paris-MVA  method~\cite{Bernlohr:2012we}.  Because the angular resolution of CTA will be small compared to the size of the ON and OFF regions and the scale on which the DM density profile varies near the region boundaries, the effect of the PSF is negligible for this study.  We note that the effective area adopted in this work was calculated for on-axis sources, and we neglect that the effective area of CTA will vary across the FOV, decreasing towards the edges; however, the regions we adopt in this analysis are close to the center of the FOV (within $2.5^\circ$) where we expect the effective area to be fairly constant.

We assume the background to be isotropic.  The number of background events observed from a specified region of the sky over an energy window $\Delta E$ is then 
\begin{equation}
\label{eq:nbg}
N_{\rm bg} = t_{\rm obs} \Delta \Omega \int_{\Delta E}{ \int_{-\infty}^{+\infty}{ \frac{dN_{\rm CRE}(\bar{E})}{dE\,dA\,dt\,d\Omega} A_{\rm eff}(\bar E) \frac{e^{-\frac{(E-\bar{E})^{2}}{2\sigma^{2}}}}{\sqrt{2\pi\sigma^{2}}} d\bar{E}}\, dE},
\end{equation}
where $\Delta \Omega$ is the solid angle of the region and $dN_{\rm CRE}/dE\,dA\,dt\,d\Omega$ is the differential intensity spectrum of the CRE events, and again we have convolved the source spectrum with the energy resolution of CTA\@.  For each DM mass, we consider energies of 30~GeV up to the DM mass in the case of annihilation, or up to half the DM mass in the case of decay, in bins of width $\Delta\log_{10}(E) = 0.15$.

For reference, assuming 200\,h of observation, an NFW profile, $m_{\chi}=1$~TeV, and annihilation to $b\bar{b}$ with $\langle \sigma v \rangle = 3 \times 10^{-26}$~cm$^{3}$~s$^{-1}$, the number of signal events in the ON region integrated from 30~GeV to 1~TeV is $N_{\rm ann, ON} \sim 2000$, while the number of signal events in the OFF region after rescaling is $\alpha N_{\rm ann, OFF} \sim 620$.  The number of background events is $N_{\rm bg,ON}=\alpha N_{\rm bg,OFF} \sim 1.5 \times 10^{6}$.

\begin{figure}[ht]
\centering
	\includegraphics[width=7cm]{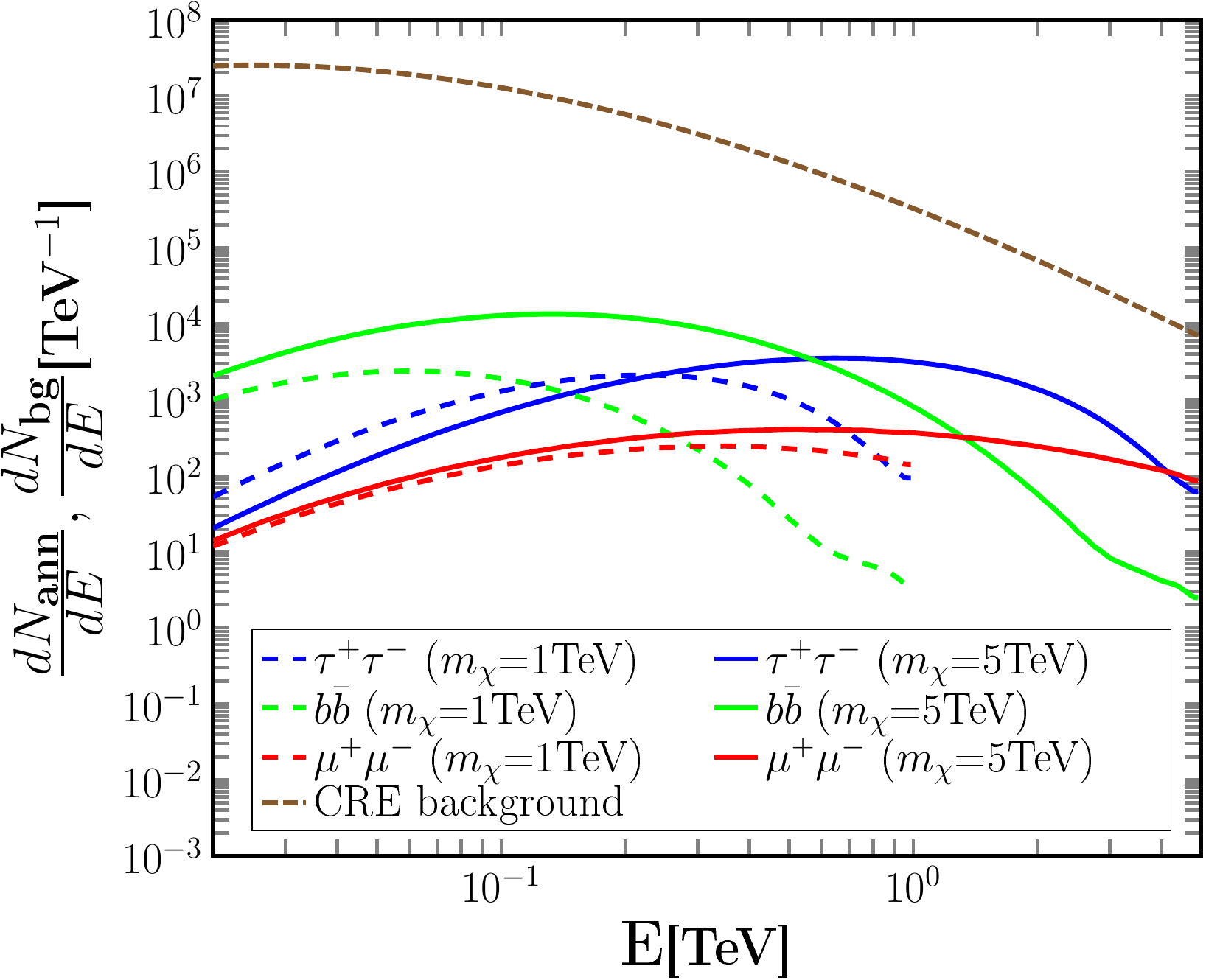}
	\caption{The observed count spectrum for annihilation to $\tau^{+}\tau^{-}$, $b\bar{b}$, and $\mu^{+}\mu^{-}$, for $m_{\chi}=1$~TeV or 5~TeV with $\langle \sigma v \rangle = 3 \times 10^{-26}$~cm$^{3}$~s$^{-1}$, for an NFW profile, compared with the CRE background.  The spectra shown are for the ON region, assume 200\,h of observation, and have been convolved with the energy resolution of CTA\@.
	\label{fig:spectra}}
\end{figure}

The observed count spectra of some example annihilation signals ($dN_{\rm ann}/dE$ with $N_{\rm ann}$ given in Eq.~\ref{eq:nann}) and the CRE background ($dN_{\rm bg}/dE$ with $N_{\rm bg}$ given in Eq.~\ref{eq:nbg}) are shown in Fig.~\ref{fig:spectra}.  The effective area of CTA increases rapidly with energy over the entire range considered here, enhancing sensitivity to high-mass DM signals.  The photon spectrum for annihilation (or decay) to $b\bar{b}$ is fairly soft and not easily distinguished from the CRE background at most energies.  In contrast, the harder spectra associated with annihilation (or decay) to $\mu^{+}\mu^{-}$ or $\tau^{+}\tau^{-}$ are distinct from the CRE background spectrum.  In all cases, the spectral cut off at the DM mass (or half the DM mass in the case of decay) provides a distinguishing feature.  

\section{Likelihood analysis}
\label{sec:likelihood}

Analyses using the Ring Method search for an excess of counts in the ON region compared to the OFF region. After rescaling the observed counts in the OFF region by the factor $\alpha$,
the excess of counts between the ON and the rescaled OFF regions is defined as  $\theta_{\rm diff}=\theta_{\rm ON}-\alpha \theta_{\rm OFF}$, where $\theta_{\rm ON}$ and $\theta_{\rm OFF}$ are the total numbers of events (the sum of signal photons and background events) in the ON and OFF regions, respectively.

Following \cite{Cline13b}, we assume that the likelihood of observing $\theta$ counts in a given region is Poisson-distributed with mean value $N$, so the likelihood of observing the difference $\theta_{\rm diff}$ is described by the Skellam distribution \cite{Skellam}:
\begin{equation}
\mathcal{L}(\theta_{\rm diff})= e^{-(N_{\rm ON}+\alpha N_{\rm OFF})}\left(\frac{N_{\rm ON}}{\alpha N_{\rm OFF}}\right)^\frac{\theta_{\rm diff}}{2} I_{|\theta_{\rm diff}|}(2\sqrt{\alpha N_{\rm ON} N_{\rm OFF}}).
\end{equation}
Here $I_{|\theta_{\rm diff}|}$ is the $|\theta_{\rm diff}|^{th}$ Bessel function of the first kind.  Note that here for simplicity we assumed a Poisson likelihood for $\alpha N_{\rm OFF}$, whereas formally $N_{\rm OFF}$ is the Poisson-distributed quantity.  The error induced by this approximation is small because $\alpha$ is not very different from 1; the approximation yields more conservative limits because it slightly over-estimates Poisson fluctuations in the OFF region.  We determine the expected limit in the case that $\theta_{\rm diff}=0$, i.e. assuming no signal and perfectly isotropic background.  The likelihood is
\begin{equation}
\mathcal{L}(m_{\chi},\pi)= e^{-(N_{\rm ON}+\alpha N_{\rm OFF})} I_{0}(2\sqrt{\alpha N_{\rm ON} N_{\rm OFF}})
\end{equation}
where $\pi$ denotes $\langle \sigma v \rangle$ for annihilation and $\tau$ for decay.

We take advantage of the spectral information (c.f.~Fig.~\ref{fig:spectra}) by calculating the likelihood over small energy bins, and define the total likelihood as the product of the likelihoods over each energy bin:
\begin{equation}
\mathcal{L}(m_{\chi},\pi)=\prod_{j}\mathcal{L}_{j}(m_{\chi},\pi),
\end{equation}
where $j$ indexes the energy bins.  We choose energy bins with log spacing of $\Delta\log_{10}(E) = 0.15$.

We calculate the likelihood ratio $-2\ln\left(\mathcal{L}(m_{\chi},\pi)/\max[\mathcal{L}(m_{\chi},\pi)]\right)$, which is $\chi^{2}$-distributed with one degree of freedom and can be well approximated by a Gaussian distribution by the central limit theorem. In the $\theta_{\rm diff}=0$ case assumed here, the 
likelihood ratio is maximized for $\langle \sigma v \rangle=0$ or $\tau=\infty$ (i.e., no signal events), so we compare this ratio to a Gaussian distribution and find the value of $\pi$, for a certain $m_{\chi}$, which constrains our model at a certain confidence level (CL).  We determine upper limits on $\langle \sigma v \rangle$ and lower limits on $\tau$ as a function of $m_{\chi}$ at 95.4\% CL.

\section{Results}
\label{sec:results}

We now present the expected sensitivity of CTA to annihilation and decay signals from various DM models. As benchmarks, we consider branching fractions of 100\% to $\tau^{+}\tau^{-}$, $b\bar{b}$, or $\mu^{+}\mu^{-}$.  Unless otherwise specified, an observation time of 200\,h is assumed, and the likelihood analysis is performed using multiple energy bins from 30~GeV to $m_{\chi}$ for annihilation (or to $m_{\chi}/2$ for decay) as described in \S\ref{sec:likelihood}.

\begin{figure}[t]
\centering
	\includegraphics[width=7cm]{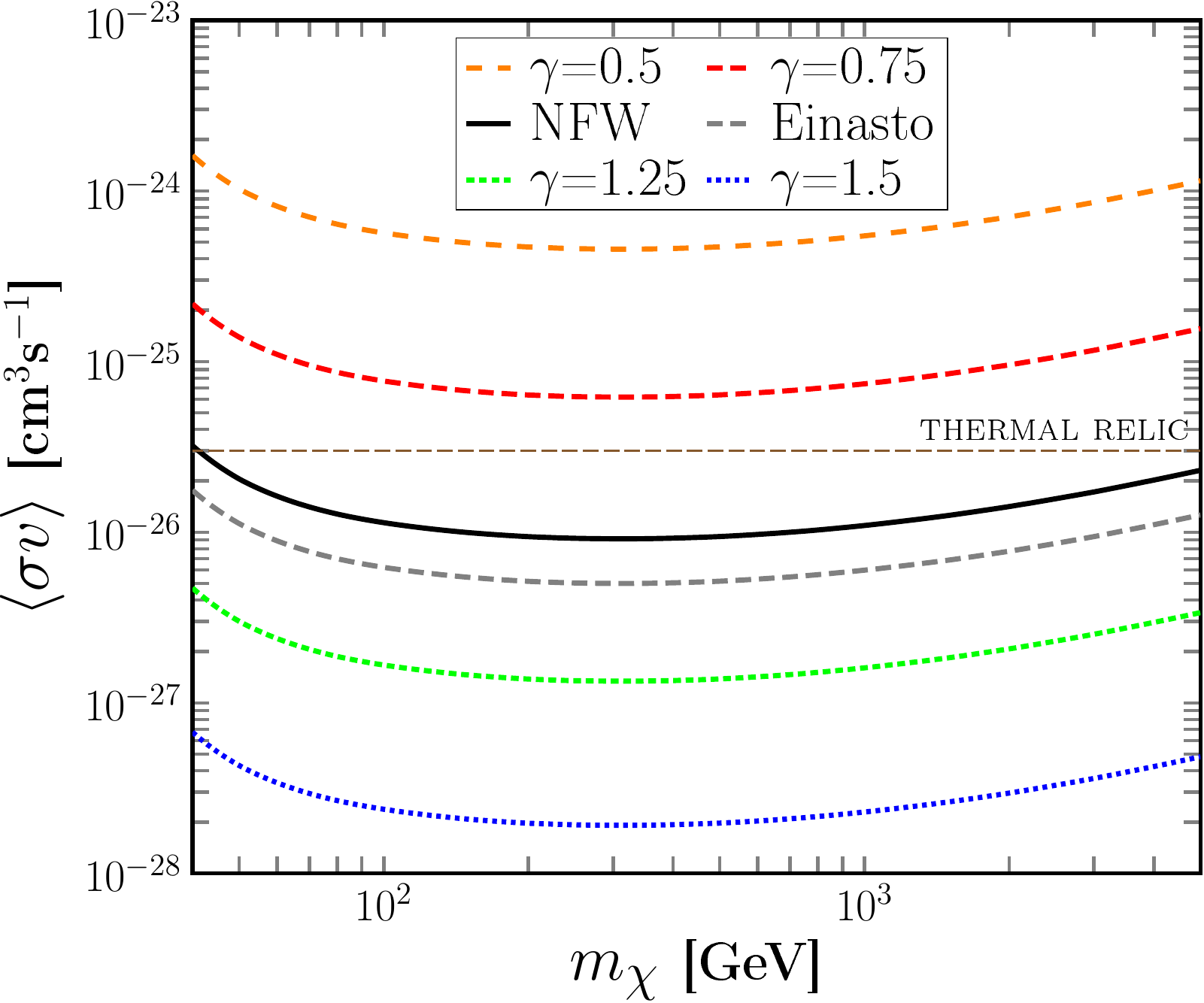}\hspace{0.5cm}
	\includegraphics[width=7cm]{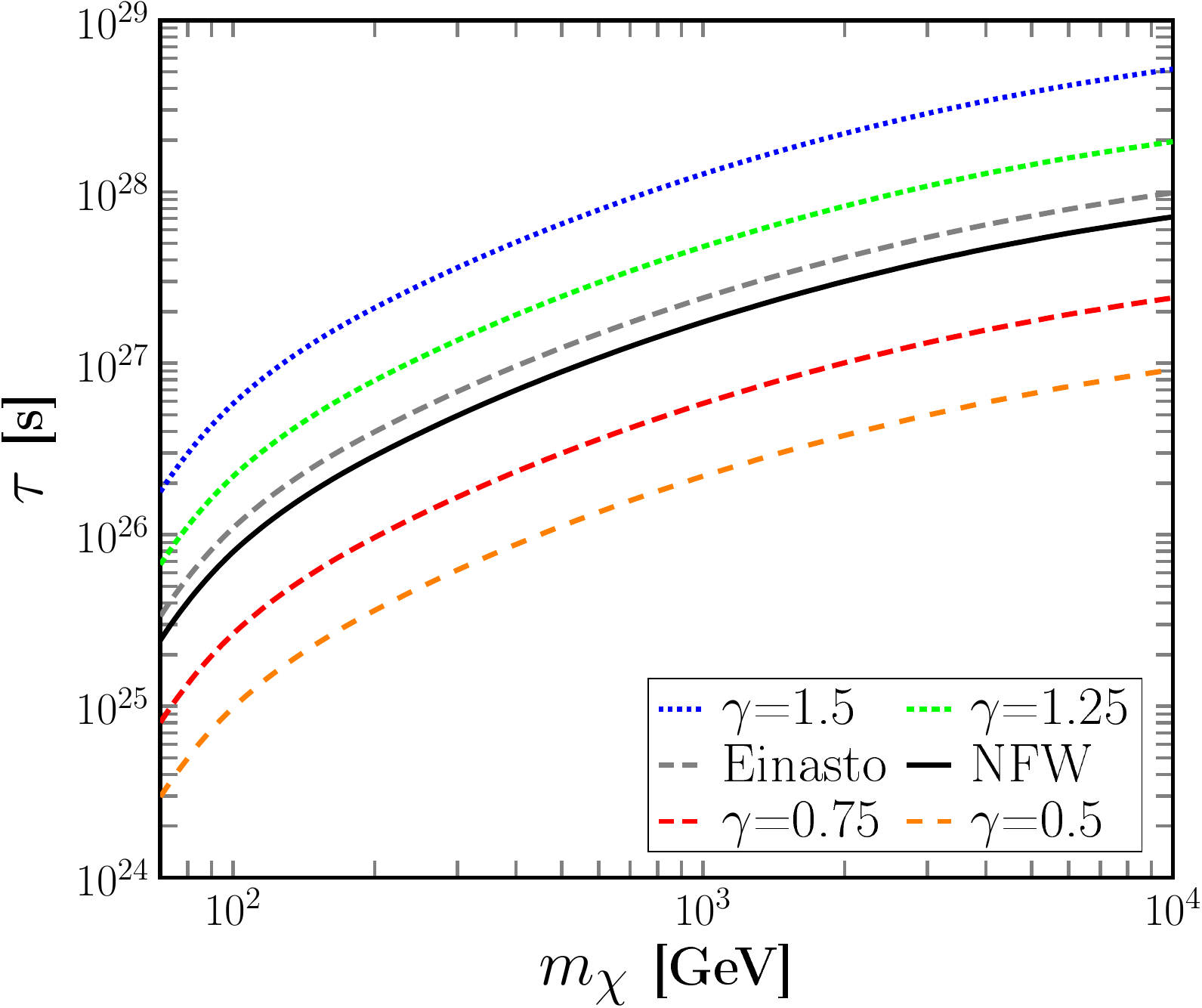}
\caption{Projected sensitivity of CTA to annihilation (\emph{left}) or decay (\emph{right}) to $\tau^{+}\tau^{-}$ for different density profiles.  Sensitivity curves are shown at 95\% CL, assuming no signal is detected (see text for details).  Curves indicate upper limit on $\langle \sigma v \rangle$ (\emph{left}) or lower limit on lifetime $\tau$ (\emph{right}).  The gray dashed line indicates the canonical value of the cross section expected if WIMPs are produced thermally with the correct relic density (although full thermal production is still viable in some scenarios with values a few orders of magnitude in either direction).
\label{fig:varprofiles}}
\end{figure}

The dependence of the sensitivity on the DM density profile is explored in Fig.~\ref{fig:varprofiles}, for annihilation or decay to $\tau^{+}\tau^{-}$.  The limits shown are at 95\% CL, assuming no difference is detected between the number of counts in the ON and rescaled OFF regions.  Varying the inner slope of the DM density profile has a substantial impact on the detectability of the signal in the case of annihilation, and a smaller but still important impact in the case of decay.  For the $\tau^{+}\tau^{-}$ channel, CTA will be sensitive to annihilation cross sections below thermal over the full range of masses probed (40~GeV to 5~TeV) for profiles with $\gamma \ge 1$ as well as the Einasto profile.  For very steep profiles $\gamma \gtrsim 1.25$, CTA will probe cross sections more than an order of magnitude below thermal.  We do not consider explicitly the case of a cored profile ($\gamma = 0$) because in this case there is essentially no variation in the signal intensity per solid angle between the ON and OFF regions, so we expect to have no sensitivity with the Ring Method analysis.  As shown in Fig.~\ref{fig:varprofiles}, with a profile as shallow as $\gamma=0.5$ the sensitivity is already quite poor.

The sensitivities we obtain for the two benchmark profiles, the NFW and Einasto profiles, are similar, with the Einasto profile providing slightly stronger sensitivity.  This is due to the overall amplitude of the signal from the Einasto profile being larger than for the NFW profile, even though the fractional difference between the ON and OFF regions is similar for both profiles.  In particular, for annihilation the $J$ factor of the Einasto profile integrated over the ON region is $\sim 7.7\times10^{21}$~GeV$^{2}$~cm$^{-5}$, whereas $\alpha$ times the integrated $J$ factor in the OFF region is $\sim 2.6\times10^{21}$~GeV$^{2}$~cm$^{-5}$; for the NFW profile the integrated ON region $J$ factor is $\sim 4.0\times10^{21}$~GeV$^{2}$~cm$^{-5}$ and $\alpha$ times the integrated OFF region $J$ factor is $\sim 1.3\times10^{21}$~GeV$^{2}$~cm$^{-5}$.

\begin{figure}[t]
\centering
	\includegraphics[width=7.cm]{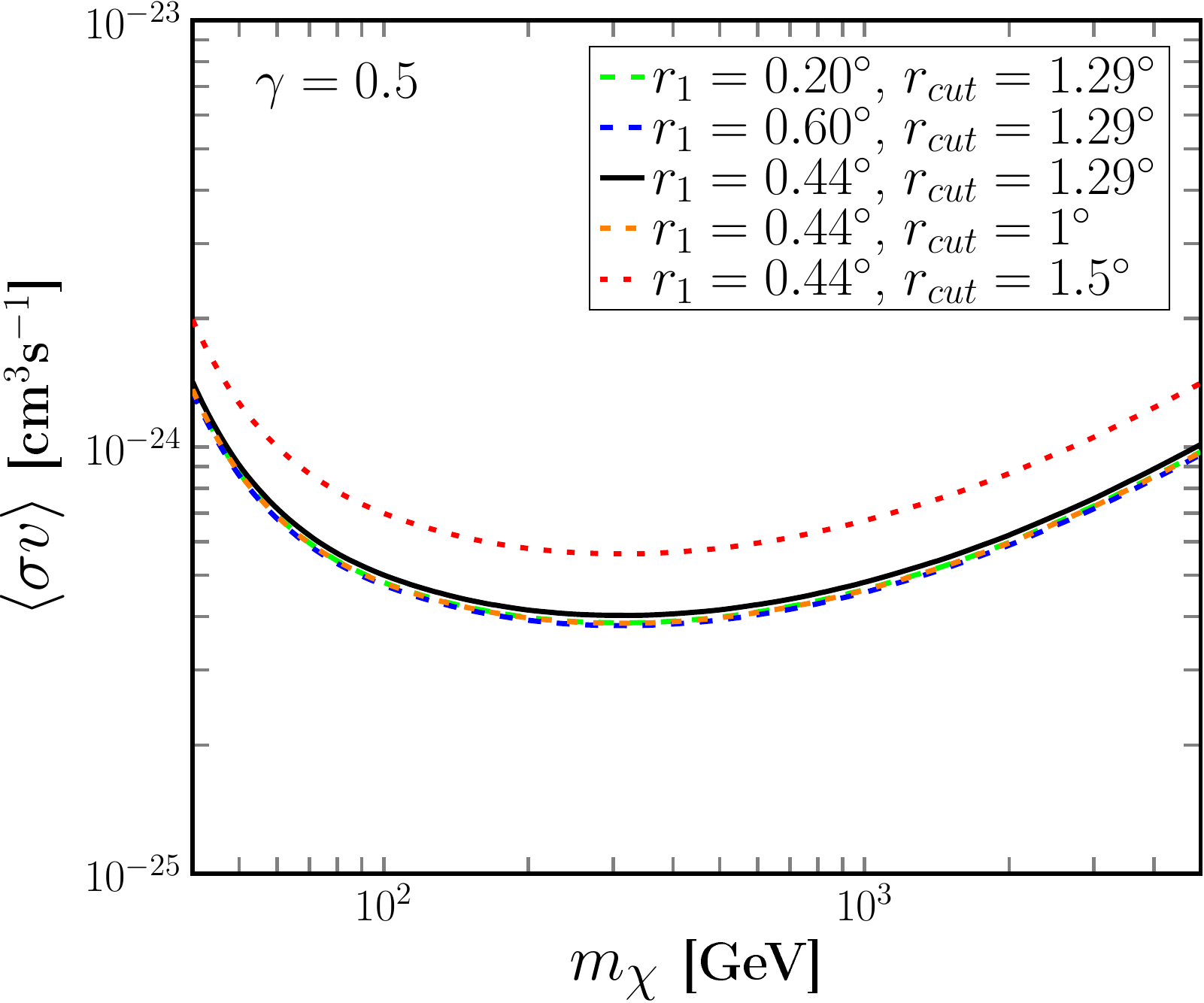}
	\hspace{.5cm}
	\includegraphics[width=7.cm]{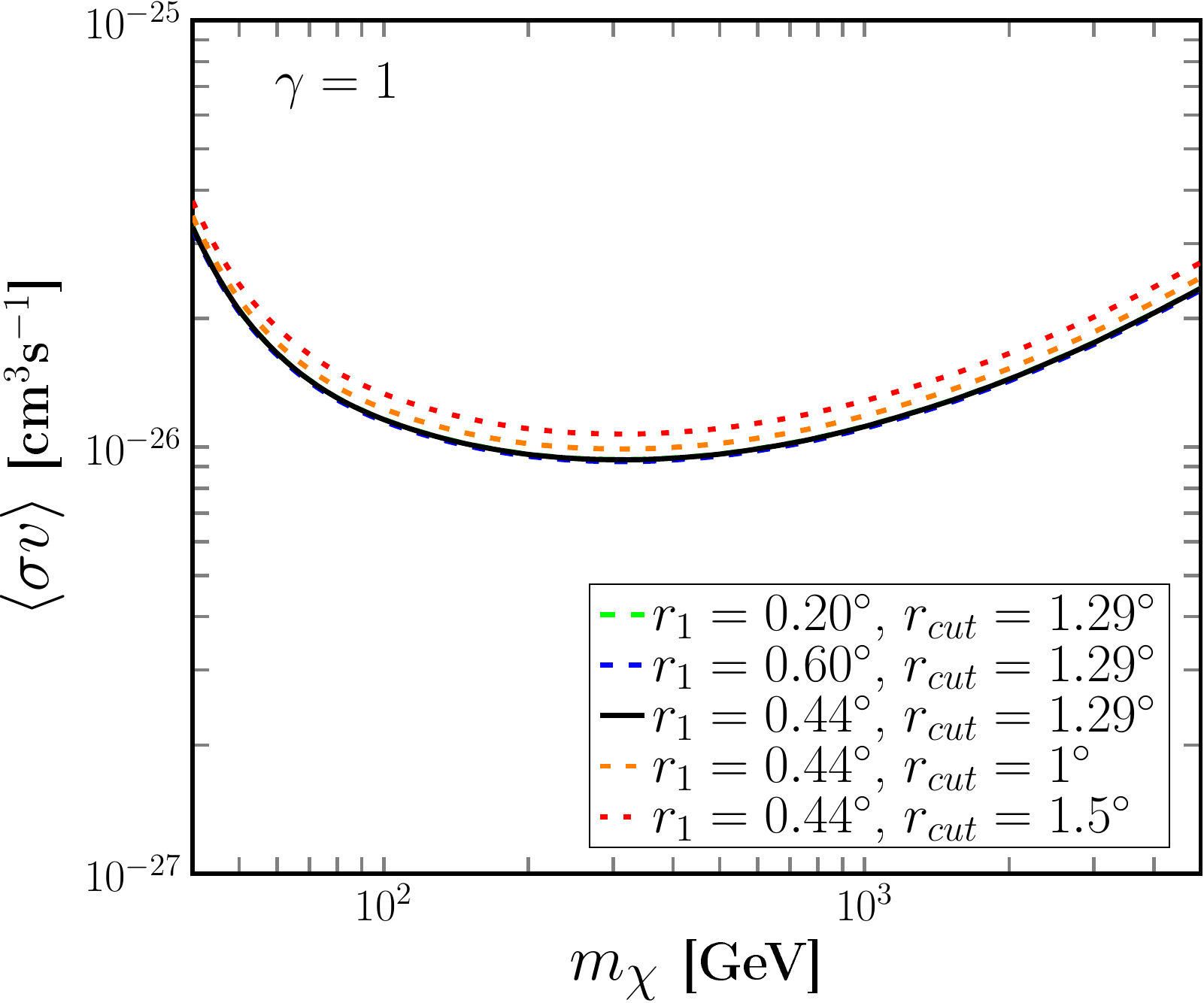}
	\includegraphics[width=7.cm]{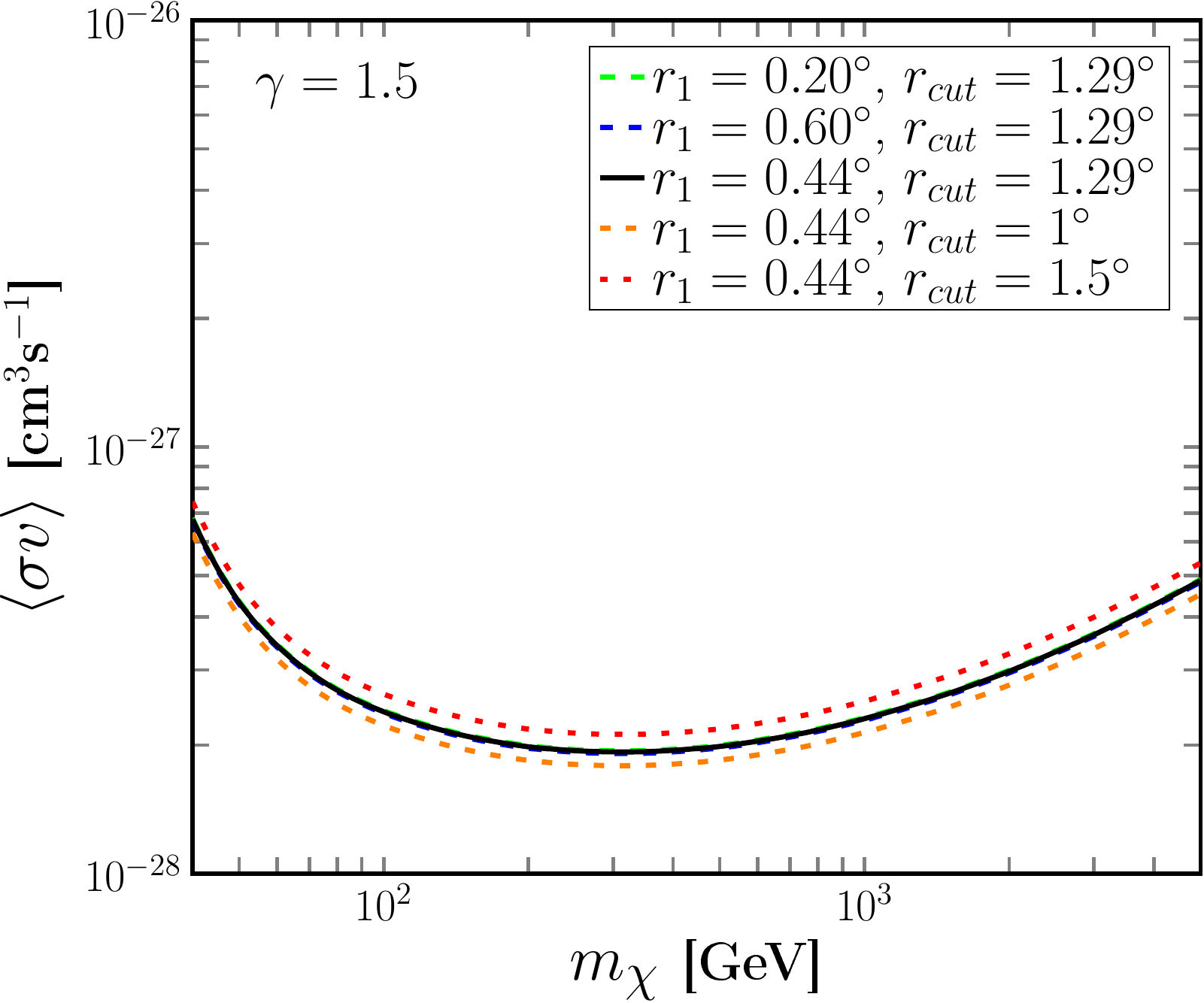}
\caption{Projected sensitivity of CTA to  
annihilation to $\tau^{+}\tau^{-}$ with varied choices of $r_{1}$ and $r_{\rm cut}$ (parameters defining the ON and OFF regions, see text for details).  The cases of $\gamma=0.5$ (\emph{top left}), $\gamma=1.0$ (\emph{top right}), and $\gamma=1.5$ (\emph{bottom}) are shown.  Upper limits on $\langle \sigma v \rangle$ are shown at 95\% CL assuming no signal is detected.  The only moderately significant change in the projected sensitivity occurs when increasing the size of the ON region via $r_{\rm cut}$ for a shallow density profile ($\gamma=0.5$).
\label{fig:ringvary}}
\end{figure}

Changing the inner slope of the DM density profile affects the integrated $J$ factors in the ON and OFF regions, so we examine whether it is possible to improve sensitivity for different density profiles by varying the geometrical parameters that define the ON and OFF regions (c.f.~Fig.~\ref{fig:ring}).  The results of varying the parameters $r_1$ (which sets the inner radius of the ring) and $r_{\rm cut}$ (which sets the radial extent of the ON region, which is centered on the GC) are shown in Fig.~\ref{fig:ringvary}.  Surprisingly, the optimal regions do not vary significantly for different density profiles.  A small improvement in sensitivity can be achieved for very steep profiles ($\gamma=1.5$) by reducing $r_{\rm cut}$.  The sensitivity weakens somewhat when increasing $r_{\rm cut}$ for a shallow density profile ($\gamma=0.5$).  Given the small degree of variation in achieved limits for relatively large changes in  $r_{\rm cut}$ and  $r_{1}$, we do not attempt to more carefully optimize the choice of ON and OFF regions, and for simplicity adopt the default parameters noted in \S\ref{sec:ring} for all cases shown.  Of course, once the final array configuration for CTA is known and characterized, optimization of the ON and OFF regions should be performed for each density profile under consideration prior to performing a DM search in the data.

\begin{figure}[t]
\centering
	\includegraphics[width=7.cm]{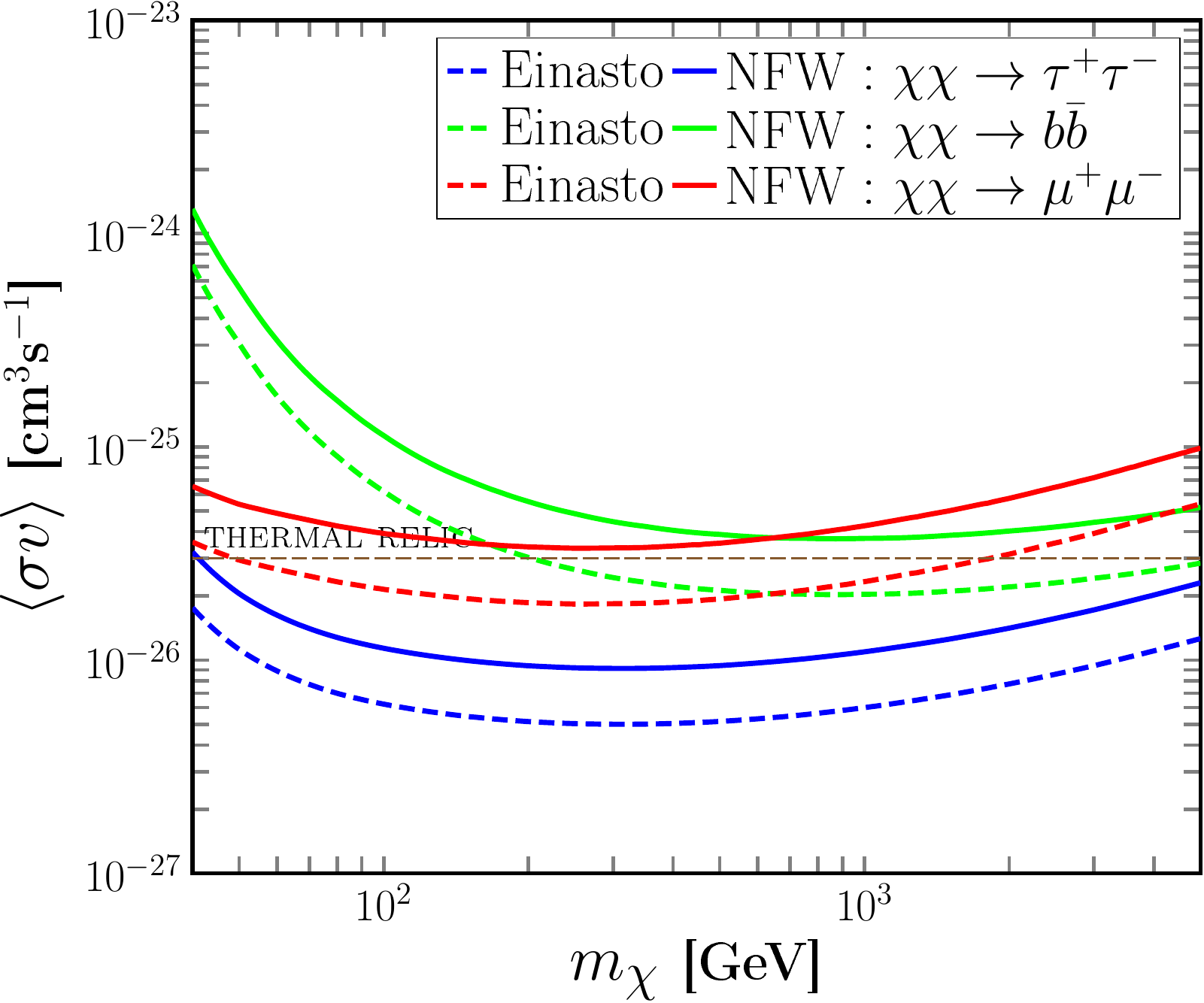}
	\hspace{.5cm}
	\includegraphics[width=7.cm]{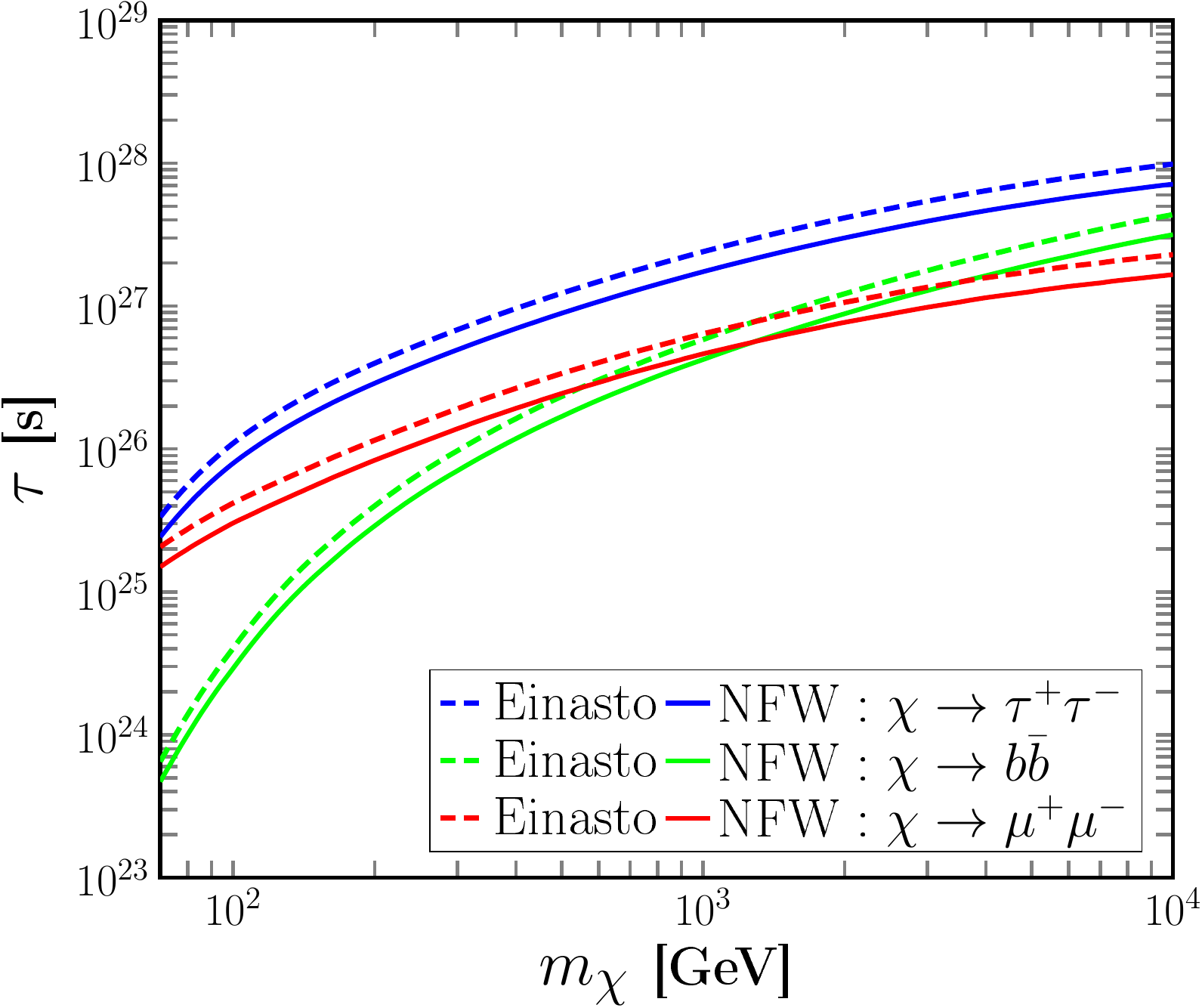}
\caption{Projected sensitivity of CTA to  
annihilation or decay to $\tau^{+}\tau^{-}$, $b\bar{b}$, and $\mu^{+}\mu^{-}$, for the NFW and Einasto profiles. \emph{Left:} For the case of annihilation, upper limits on $\langle \sigma v \rangle$ at 95\% CL if no signal is detected.    The gray dashed line is the canonical thermal cross section, but full thermal production can still be viable with cross-sections a few orders of magnitude higher or lower.  \emph{Right:} Lower limits on lifetime $\tau$ at 95\% CL if no signal is detected, for the case of decay.
\label{fig:channels}}
\end{figure}

The projected sensitivity of CTA to benchmark annihilation and decay channels is shown in Fig.~\ref{fig:channels}, for the NFW and Einasto density profiles.   
The sensitivity for the $b\bar{b}$ final state is weaker at low DM masses due to the softness of the spectrum resulting in fewer signal events above the energy threshold of 30~GeV.
Note that for a given annihilation or decay channel, the relative sensitivity for other density profiles scales in the same way as in Fig.~\ref{fig:varprofiles}.

\begin{figure}[t]
\centering
	\includegraphics[width=7cm]{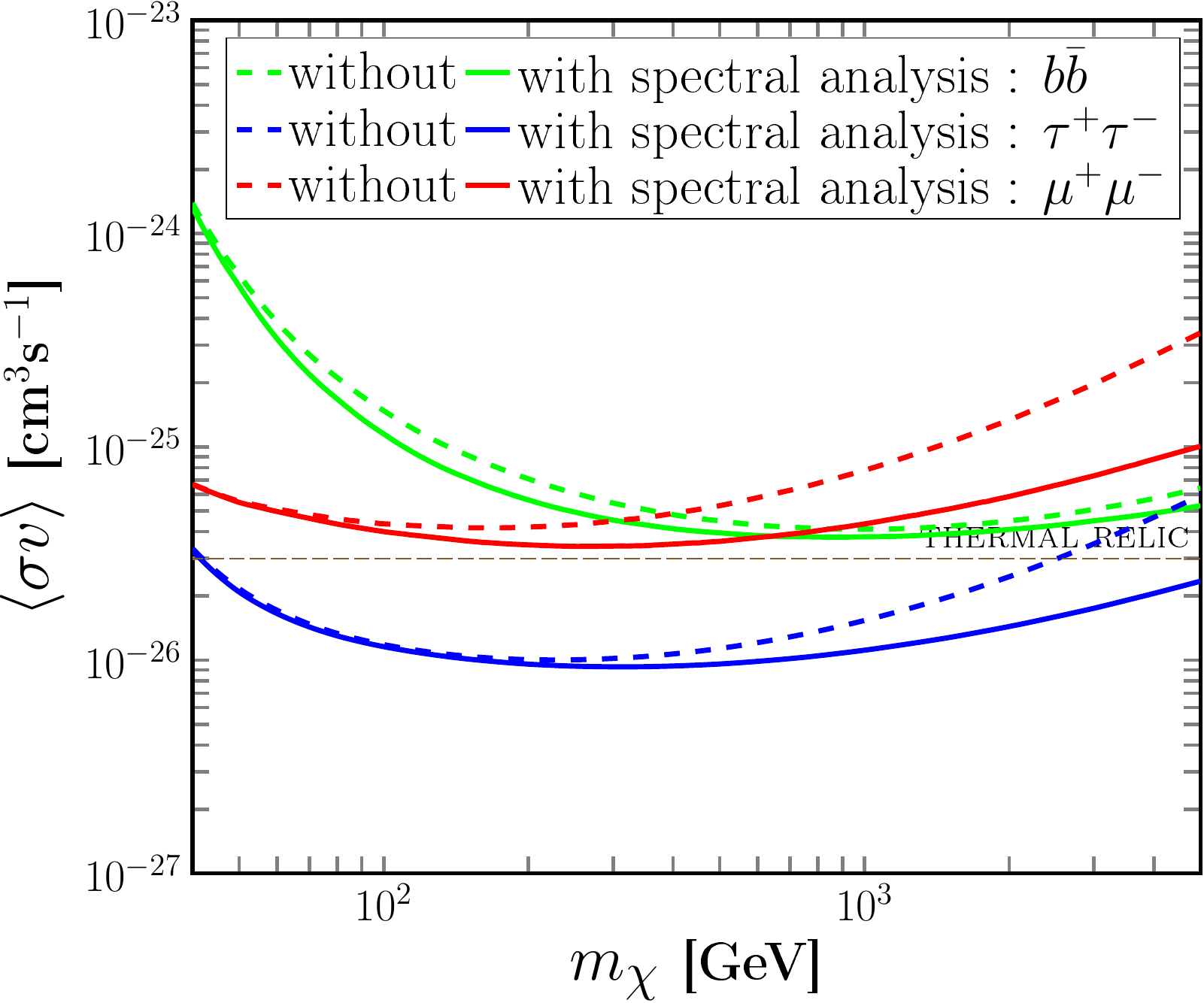}\hspace{0.5cm}
	\includegraphics[width=7cm]{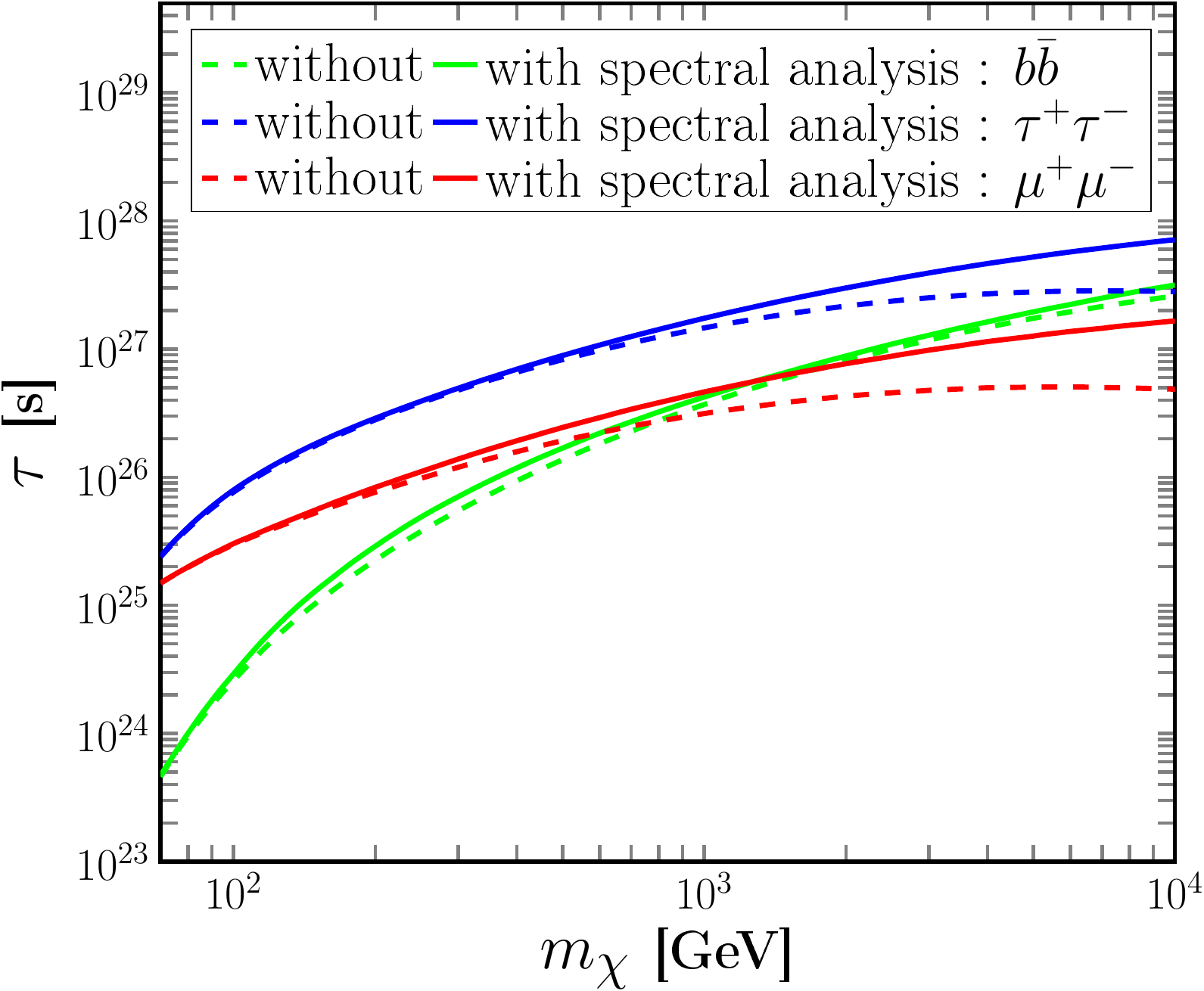}
\caption{Impact of spectral analysis on sensitivity.  Sensitivity curves at 95\% CL are shown for annihilation (\emph{left}) or decay (\emph{right}) to $b\bar{b}$ (\emph{green}), $\tau^{+}\tau^{-}$ (\emph{blue}), or $\mu^{+}\mu^{-}$ (\emph{red}) for an NFW profile, with (\emph{solid}) or without (\emph{dashed}) performing the likelihood analysis over multiple energy bins.  See text for details.  The gray dashed line is the canonical thermal annihilation cross section, but full thermal production can still be viable with cross-sections a few orders of magnitude higher or lower.
\label{fig:spectanalysis}}
\end{figure}

We show the impact of performing a spectral analysis in Fig.~\ref{fig:spectanalysis}.  Here we show the sensitivity for an NFW profile, either integrating the observed events over one energy bin, or utilizing multiple energy bins as we describe in \S\ref{sec:likelihood}.  The effect is strongest for the $\mu^{+}\mu^{-}$ and $\tau^{+}\tau^{-}$ channels at moderate to high DM masses, and improves the expected sensitivity for annihilation or decay by up to a factor of $\sim 3$.

\begin{figure}[t]
\centering
	\includegraphics[width=7cm]{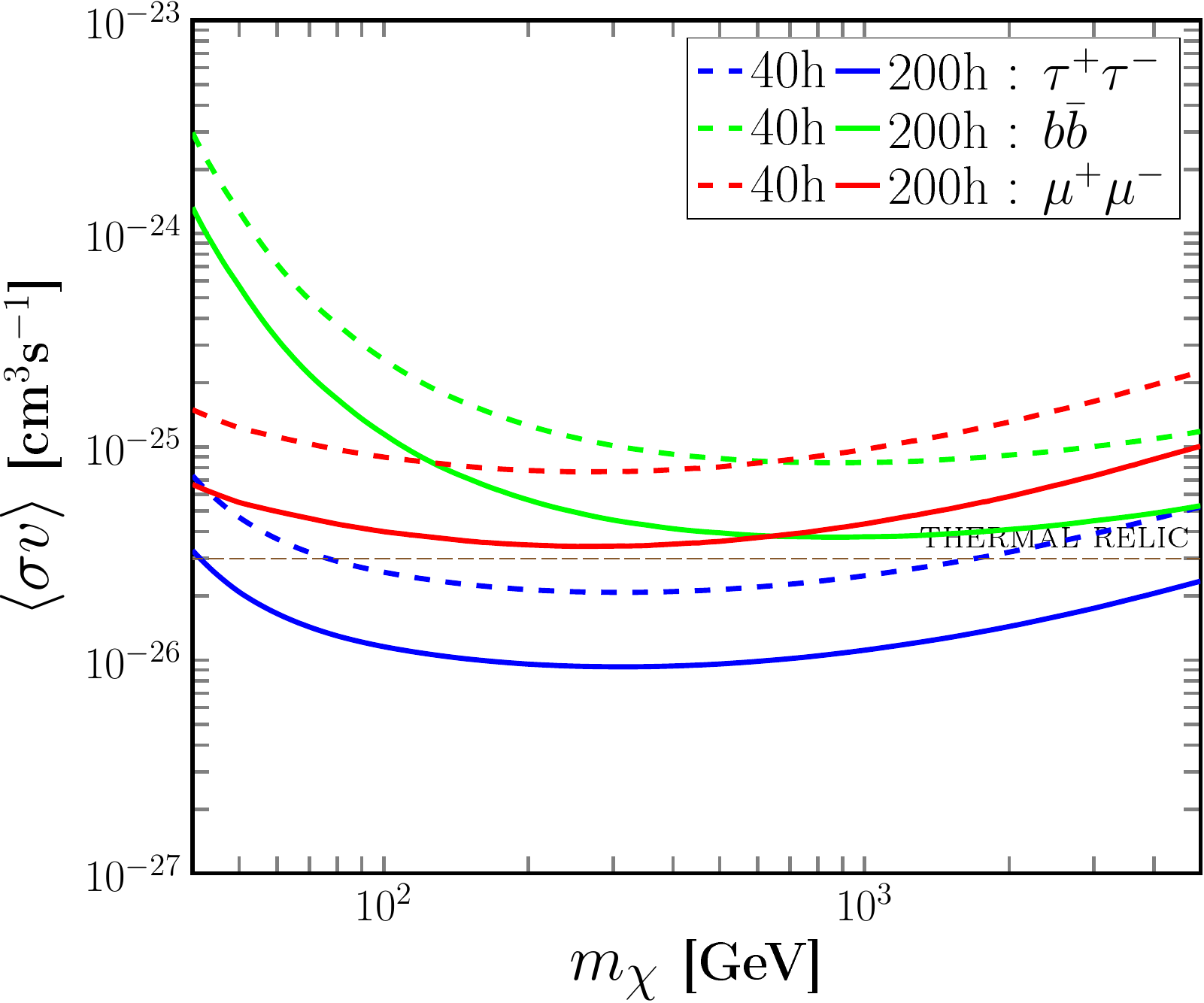}\hspace{0.5cm}
	\includegraphics[width=7cm]{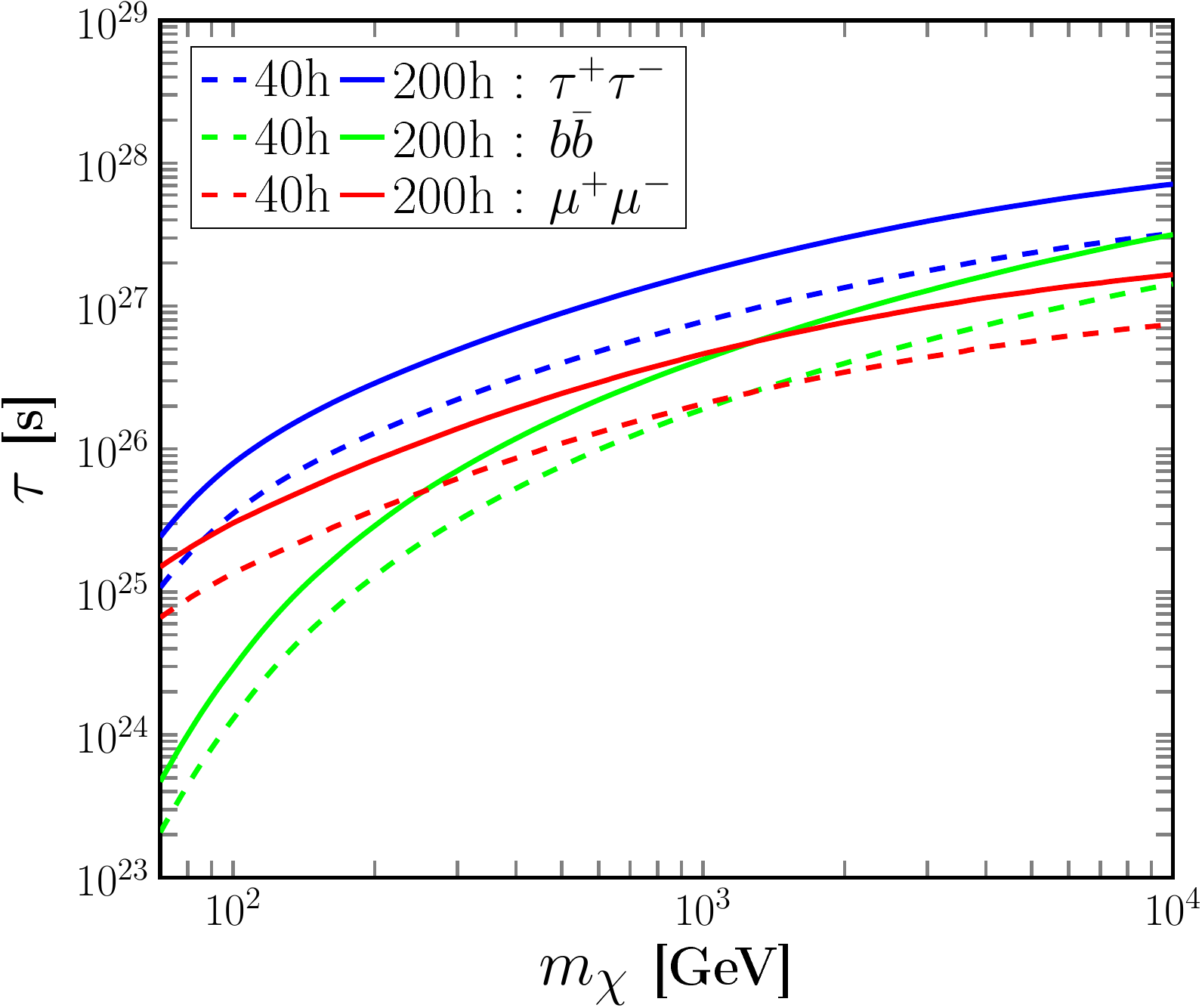}
\caption{Dependence of sensitivity on observation time.  Sensitivity curves at 95\% CL are shown for annihilation (\emph{left}) or decay (\emph{right}) to $\tau^{+}\tau^{-}$ (\emph{blue}), $b\bar{b}$ (\emph{green}), or $\mu^{+}\mu^{-}$ (\emph{red}) for an NFW profile, assuming 40\,h (\emph{dashed}) or 200\,h (\emph{solid}) of observation time.  The gray dashed line is the canonical thermal annihilation cross section, but full thermal production can still be viable with cross-sections a few orders of magnitude higher or lower.
\label{fig:tobscompare}}
\end{figure}

Figure~\ref{fig:tobscompare} demonstrates the dependence of the sensitivity on observation time assuming an NFW profile.  The sensitivity at all masses for both annihilation and decay to $b\bar{b}$, $\tau^{+}\tau^{-}$, and $\mu^{+}\mu^{-}$ improves roughly as the square root of the observation time, indicating that the expected limit is background-dominated.  However, for annihilation to $\tau^{+}\tau^{-}$, even 40\,h of observation will be sufficient to probe the canonical thermal cross section for $m_{\chi}$ from $\sim 80$~GeV to $\sim 2$~TeV for an NFW profile.

\begin{figure}[t]
\centering
\includegraphics[width=8cm]{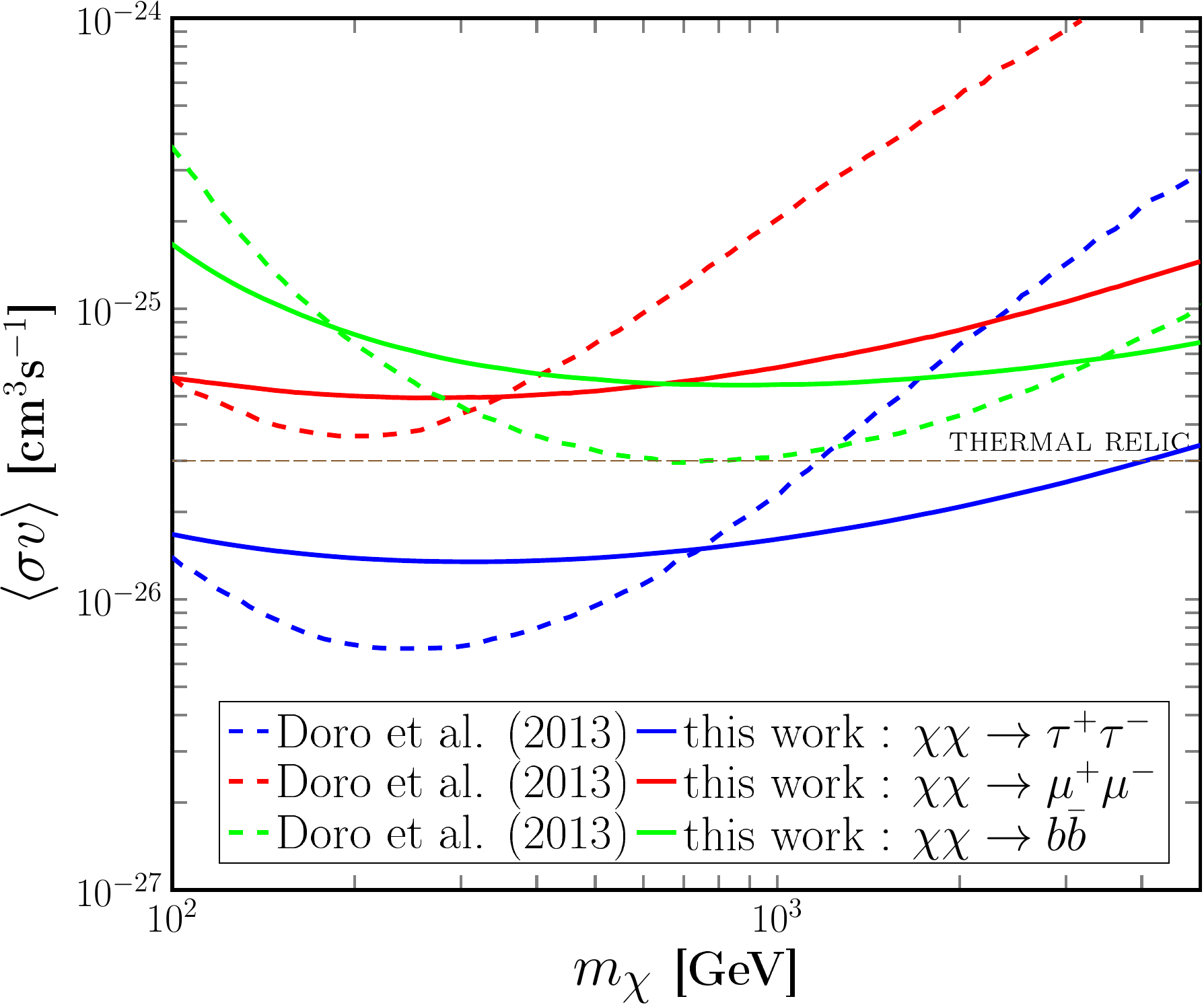}
\caption{Comparison of projected sensitivity of CTA from this work (\emph{solid curves}), which uses Array I, to the CTA Consortium estimate from~\cite{Doro:2012xx} for Array E (\emph{dashed curves}), assuming an NFW profile and 100\,h of observation. Projected sensitivities are calculated for 3 different annihilation channels: $\tau^{+}\tau^{-}$ (\emph{blue}), $b\bar{b}$ (\emph{green}), and $\mu^{+}\mu^{-}$ (\emph{red}).  Differences in projected sensitivity arise due to different assumptions regarding the array configuration and differences in the analysis method; see text for details.
\label{fig:CTAcompare}}
\end{figure}

We now compare in Fig.~\ref{fig:CTAcompare} our projected sensitivity with the sensitivity determined by the CTA Consortium~\cite{Doro:2012xx}.  In this figure the curves correspond to 100\,h of observation, and we have adopted the NFW profile.
The results of~\cite{Doro:2012xx} are stronger by at most a factor of a few, with the largest improvements for $m_{\chi}$ of $\sim 200$~GeV for $\tau^{+}\tau^{-}$ and $\mu^{+}\mu^{-}$, and for $m_{\chi}$ of $\sim 700$~GeV for $b\bar{b}$. For higher DM masses ($\sim$~few TeV), our derived sensitivity is stronger by a factor of $\sim 10$ for $\tau^{+}\tau^{-}$ and $\mu^{+}\mu^{-}$. 

Several differences in the assumptions and analysis methods employed in our work and that of \cite{Doro:2012xx} are responsible for these discrepancies.  We assumed Array I, whereas the results shown in Fig.~\ref{fig:CTAcompare} from~\cite{Doro:2012xx} assume Array E\@.  Array E includes more large telescopes, so has better sensitivity at low energies, leading to the stronger projected DM sensitivity of~\cite{Doro:2012xx} at low DM masses for all channels.  Indeed, differences in the energy-dependence of the effective areas of the two arrays leads to general differences between the two results, over the whole range of DM masses.  We also performed spectral analysis, which improves our estimated sensitivity at high DM masses relative to the results of \cite{Doro:2012xx}.  Comparing with Fig.~\ref{fig:spectanalysis}, we see that the improvement in our analysis relative to theirs at high masses is partially attributable to the spectral analysis, and partially due to the different energy-dependences of the Array I and E effective areas.  We also neglected hadronic backgrounds, which may have led to slightly optimistic results in our analysis for very low and very high DM masses.

\begin{figure}[t]
	\centering
	\includegraphics[width=7cm]{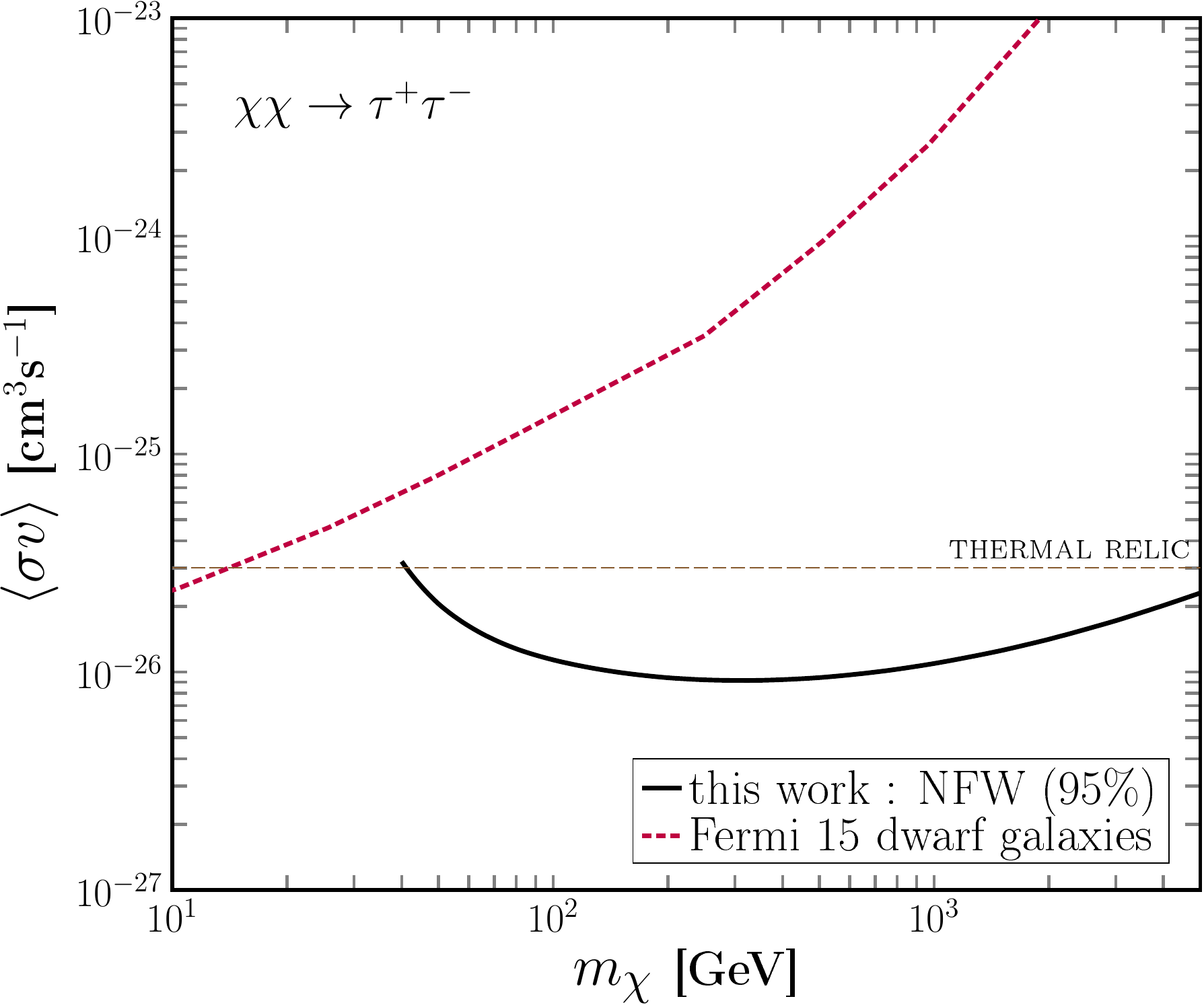}\hspace{0.5cm}
	\includegraphics[width=7cm]{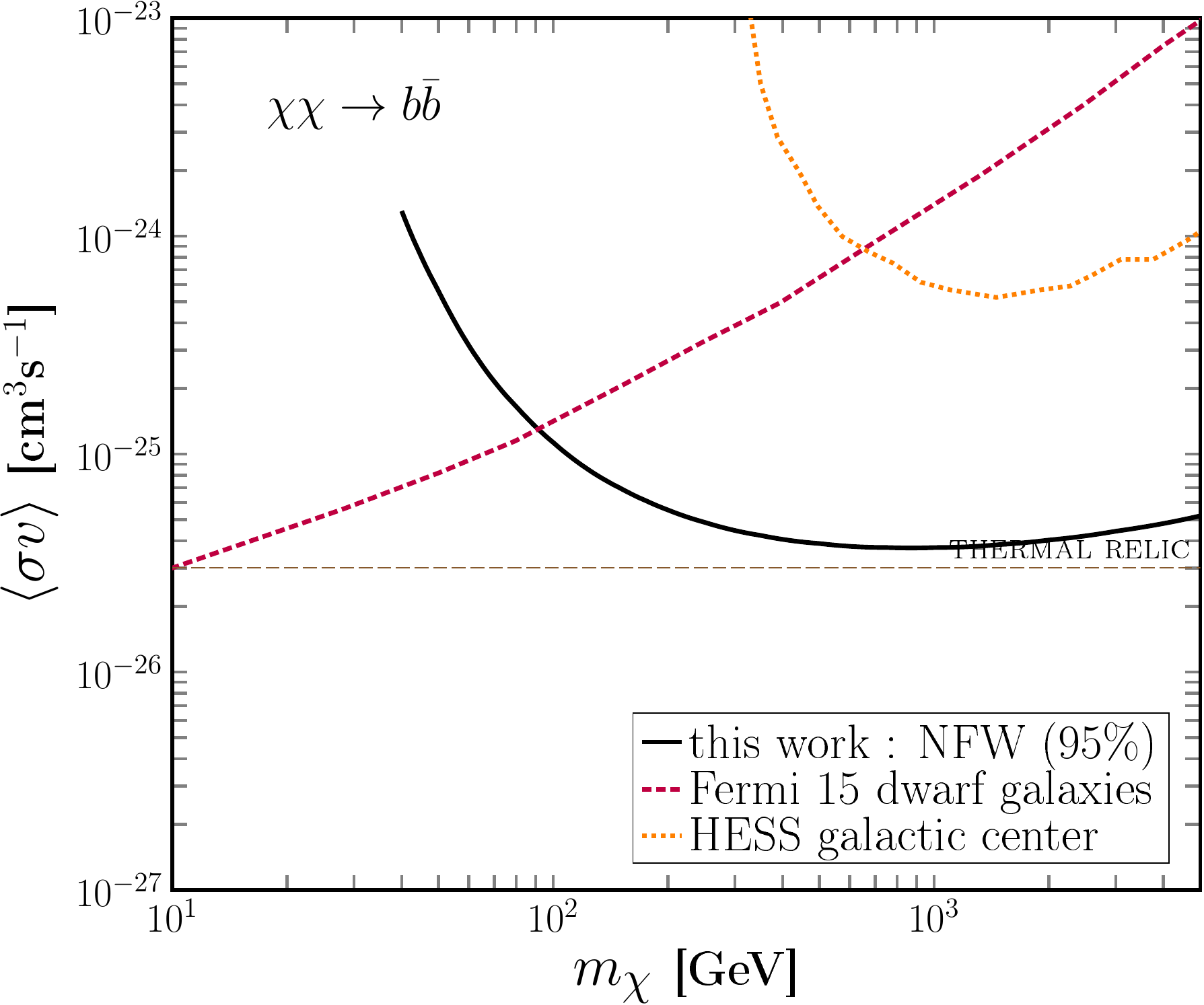}
\caption{Comparison of projected sensitivity of CTA from this work for the NFW profile (\emph{solid black curve}) to current bounds on annihilation to $\tau^{+}\tau^{-}$ (\emph{left}) and $b\bar{b}$ (\emph{right}).  The upper limit on the annihilation cross section from the Fermi LAT combined analysis of dwarf galaxies~\cite{Ackermann:2013yva} (\emph{dotted red curve}) is shown for comparison. For the $b\bar{b}$ channel the upper limit from the H.E.S.S.~GC analysis assuming an NFW profile~\cite{Abramowski:2011hc} is also shown.  The gray dashed line is the canonical thermal cross section, but full thermal production can still be viable with cross-sections a few orders of magnitude higher or lower.
\label{fig:taubbcompare}}
\end{figure}

In Fig.~\ref{fig:taubbcompare}, we show our results for annihilation to $\tau^{+}\tau^{-}$ and $b\bar{b}$, along with current constraints on these channels from the Fermi LAT combined analysis of dwarf spheroidal galaxies~\cite{Ackermann:2013yva}.  For annihilation to $b\bar{b}$ we also show the upper limit obtained from H.E.S.S.~observations of the GC, assuming an NFW profile~\cite{Abramowski:2011hc}.  To facilitate comparison, the H.E.S.S.~curve has been rescaled to correspond to a density profile with the local DM density assumed in this work of $\rho_{\odot}=.43$~GeV/cm$^{3}$; the original H.E.S.S.~result assumed a local DM density of $\rho_{\odot}=.39$~GeV/cm$^{3}$.  It is clear that a GC search with CTA will explore complementary parameter space to that probed by the Fermi LAT, and will improve on the existing IACT bound from H.E.S.S.~substantially by improving sensitivity by more than an order of magnitude in annihilation cross section and by extending the search to much lower DM masses than accessible with the H.E.S.S. observations.

\begin{figure}[t]
\centering
\includegraphics[width=8cm]{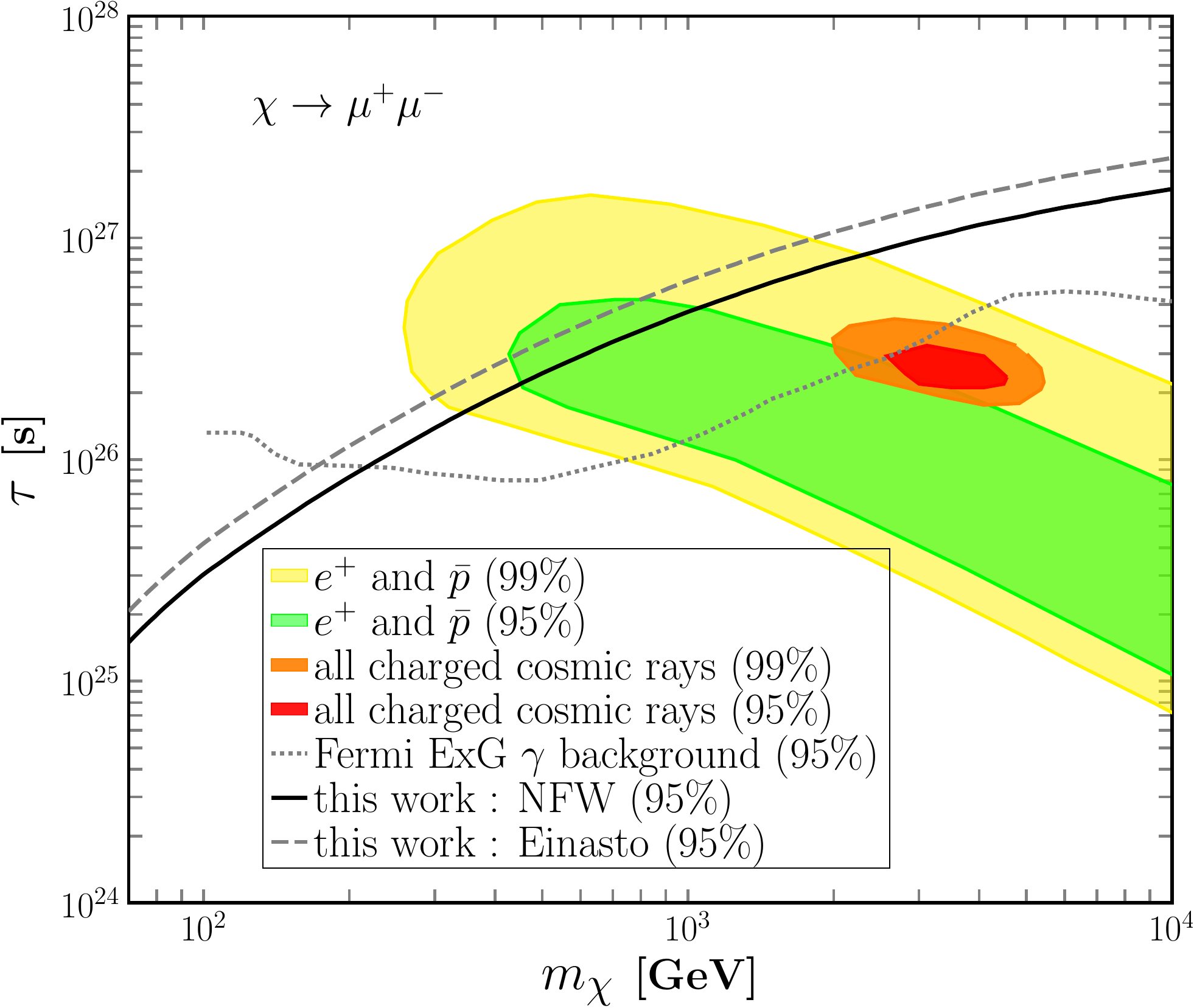}
\caption{Predicted sensitivity of CTA to DM decay to $\mu^{+}\mu^{-}$ as derived in this work, assuming an NFW or Einasto profile.  Curves indicate current lower limit or projected limit in the case of no detection on DM lifetime at 95\% CL.  Current constraints from the Fermi LAT measurement of the IGRB~\cite{Abdo:2010nz} as derived in~\cite{Cirelli:2012ut} are shown for comparison.  Regions consistent with the positron fraction measurements by PAMELA and Fermi and the PAMELA measurement of the antiproton flux, at 95.45\% C.L. and 99.999\% C.L. (green and yellow regions, respectively), are marked.  Regions consistent when the $e^{+}+e^{-}$ fluxes measured by Fermi, H.E.S.S., and MAGIC are included in the fit are also shown at 95.45\% C.L. and 99.999\% C.L. (red and orange regions, respectively)~\cite{Cirelli:2012ut}.
\label{fig:decaycompare}
}
\end{figure}

We examine the projected sensitivity of CTA to DM decay in the context of apparent anomalies observed in cosmic-ray spectra.  Figure~\ref{fig:decaycompare} shows the expected sensitivity of CTA to decay to $\mu^{+}\mu^{-}$, compared with the regions consistent with a DM explanation of the excesses in the positron fraction and total CRE spectrum measured by various experiments and consistent with the measured antiproton spectrum, as determined in~\cite{Cirelli:2012ut}.  The lower bound on the DM lifetime from the Fermi LAT measurement of the IGRB from Fig.~2 of~\cite{Cirelli:2012ut} is also shown.  The IGRB curve corresponds to constraints from the 2010 data only, excluding the later preliminary IGRB data points.  The ``signal'' regions consistent with DM interpretations of the cosmic-ray anomalies have been rescaled so that they correspond to a Milky Way halo with the local DM density adopted throughout this work.  Constraints in~\cite{Cirelli:2012ut} were derived for an NFW density profile, however the impact on the local cosmic-ray fluxes of varying the density profile is small. While currently some parameter space consistent with all of the charged cosmic-ray data remains viable for the $\mu^{+}\mu^{-}$ channel, we find that CTA will be able to test these signal regions for either an NFW or Einasto density profile.  We note  that CTA is projected to achieve similar sensitivity to a DM decay signal from Fornax~\cite{Cirelli:2012ut}.

Finally, we point out that our analyses here (and all those performed to date), have assumed isotropic backgrounds and no systematic errors, so should still be considered somewhat optimistic.  In reality the CRE and cosmic-ray proton backgrounds may not be perfectly isotropic, which can introduce a theoretical error in the calculation of $\theta_{\rm diff}$.  A far larger systematic probably comes from the Monte Carlo determination of the effective area itself.  Another source of systematic error is the modeling of the background spectra: although the slope and normalization is rather well determined for CREs, some freedom exists in the choice of CR proton parameters.  Determining the impacts of these systematics is beyond the scope of the current paper, but they should be investigated in detail if robust results are to be gleaned from CTA\@.

\section{Conclusions}
\label{sec:conc}

In this work we examined the sensitivity of searches for gamma rays from DM annihilation or decay from the GC with CTA\@.  We showed that the sensitivity of a search is strongly affected by the assumed DM density profile.  This is due both to the variation in the overall amplitude of the DM flux and to the ON-OFF analysis method adopted to determine the level of the cosmic-ray background.  If DM searches with CTA find no detection, the strong dependence of the inferred limit on the assumed density profile indicates that independent constraints on the DM density profile will be extremely important to enable meaningful comparisons between the results of this search and complementary searches, e.g., direct and collider searches.

For annihilation to $\tau^{+}\tau^{-}$, 200\,h of observation of the GC with CTA will be able to probe annihilation cross sections below the canonical thermal relic value for DM masses from $\sim 40$~GeV up to several TeV; for annihilation to $b\bar{b}$ these observations could test close to the thermal relic cross section for masses above a few hundred GeV.  The projected sensitivities for both of these channels represent a substantial improvement over existing bounds, and will allow favored models to be tested in this mass range for the first time.  CTA will also be able to confirm or exclude currently-allowed interpretations of measured cosmic-ray excesses in terms of DM decay.  Taking advantage of the differences in the spectra of the signal and background can improve sensitivity by a factor of $\sim 3$ at high DM masses for both annihilation and decay for the $\tau^{+}\tau^{-}$ and $\mu^{+}\mu^{-}$ channels.

CTA will provide the high-energy astrophysics community with a leap in observational capabilities, and has the potential to make great strides in exploring new DM parameter space through searches for annihilation and decay signals.  Upcoming cosmic-ray measurements and multi-wavelength analysis can also serve as valuable tests of DM signals, and are complementary to gamma-ray studies.  In conjunction with these other indirect searches, CTA is a promising tool to detect and constrain the particle properties of DM\@.

\acknowledgments
It is a pleasure to thank Jan Conrad, Michele Doro, Christian Farnier, Emmanuel Moulin, Rene Ong, Massimo Persic, Miguel \'{A}ngel S\'{a}nchez-Conde, Louie Strigari, Vladimir Vassiliev, Christoph Weniger, and Matthew Wood for helpful discussions.  MP acknowledges the Moore Center for Theoretical Cosmology and Physics for support and Caltech for hospitality.  JSG acknowledges support from NASA through Einstein Postdoctoral Fellowship grant PF1-120089 awarded by the Chandra X-ray Center, which is operated by the Smithsonian Astrophysical Observatory for NASA under contract NAS8-03060. PS is supported by the Banting Fellowship, administered by NSERC for the Canadian Tri-Agency Research Councils. JSG and PS thank the Kavli Institute for Theoretical Physics for hospitality.

\bibliography{CTA_GC}

\providecommand{\href}[2]{#2}\begingroup\raggedright\begin{thebibliography}{10}

\bibitem{Bertone:2004pz}
G.~Bertone, D.~Hooper, and J.~Silk, {\it {Particle dark matter: Evidence,
  candidates and constraints}},  {\em Phys.Rept.} {\bf 405} (2005) 279--390,
  [\href{http://xxx.lanl.gov/abs/hep-ph/0404175}{{\tt hep-ph/0404175}}].

\bibitem{Ade:2013zuv}
Planck Collaboration: P.~Ade {\em et.~al.}, {\it {Planck 2013 results. XVI.
  Cosmological parameters}},  \href{http://xxx.lanl.gov/abs/1303.5076}{{\tt
  arXiv:1303.5076}}.

\bibitem{BringmannWeniger}
T.~{Bringmann} and C.~{Weniger}, {\it {Gamma ray signals from dark matter:
  Concepts, status and prospects}},  {\em Physics of the Dark Universe} {\bf 1}
  (2012) 194--217, [\href{http://xxx.lanl.gov/abs/1208.5481}{{\tt
  arXiv:1208.5481}}].

\bibitem{Jungman:1995df}
G.~Jungman, M.~Kamionkowski, and K.~Griest, {\it {Supersymmetric dark matter}},
   {\em Phys.Rept.} {\bf 267} (1996) 195--373,
  [\href{http://xxx.lanl.gov/abs/hep-ph/9506380}{{\tt hep-ph/9506380}}].

\bibitem{Steigman:2012nb}
G.~Steigman, B.~Dasgupta, and J.~F. Beacom, {\it {Precise Relic WIMP Abundance
  and its Impact on Searches for Dark Matter Annihilation}},  {\em Phys.Rev.}
  {\bf D86} (2012) 023506, [\href{http://xxx.lanl.gov/abs/1204.3622}{{\tt
  arXiv:1204.3622}}].

\bibitem{Ackermann:2013yva}
Fermi-LAT Collaboration: M.~Ackermann {\em et.~al.}, {\it {Dark Matter
  Constraints from Observations of 25 Milky Way Satellite Galaxies with the
  Fermi Large Area Telescope}},  \href{http://xxx.lanl.gov/abs/1310.0828}{{\tt
  arXiv:1310.0828}}.

\bibitem{Hooper:2012sr}
D.~Hooper, C.~Kelso, and F.~S. Queiroz, {\it {Stringent and Robust Constraints
  on the Dark Matter Annihilation Cross Section From the Region of the Galactic
  Center}},  {\em Astropart.Phys.} {\bf 46} (2013) 55--70,
  [\href{http://xxx.lanl.gov/abs/1209.3015}{{\tt arXiv:1209.3015}}].

\bibitem{Bergstrom:1997fj}
L.~Bergstrom, P.~Ullio, and J.~H. Buckley, {\it {Observability of gamma-rays
  from dark matter neutralino annihilations in the Milky Way halo}},  {\em
  Astropart.Phys.} {\bf 9} (1998) 137--162,
  [\href{http://xxx.lanl.gov/abs/astro-ph/9712318}{{\tt astro-ph/9712318}}].

\bibitem{Gomez-Vargas:2013bea}
G.~A. Gomez-Vargas, M.~A. Sanchez-Conde, {\em et.~al.}, {\it {Constraints on
  WIMP Annihilation for Contracted Dark Matter in the Inner Galaxy with the
  Fermi-LAT}},  \href{http://xxx.lanl.gov/abs/1308.3515}{{\tt
  arXiv:1308.3515}}.

\bibitem{Navarro:1995iw}
J.~F. Navarro, C.~S. Frenk, and S.~D. White, {\it {The Structure of cold dark
  matter halos}},  {\em Astrophys.J.} {\bf 462} (1996) 563--575,
  [\href{http://xxx.lanl.gov/abs/astro-ph/9508025}{{\tt astro-ph/9508025}}].

\bibitem{1965TrAlm...5...87E}
J.~{Einasto}, {\it {On the Construction of a Composite Model for the Galaxy and
  on the Determination of the System of Galactic Parameters}},  {\em Trudy
  Astrofizicheskogo Instituta Alma-Ata} {\bf 5} (1965) 87--100.

\bibitem{nfwsmooth}
J.~F. {Navarro}, E.~{Hayashi}, {\em et.~al.}, {\it {The inner structure of
  {$\Lambda$}CDM haloes - III. Universality and asymptotic slopes}},  {\em
  \mnras} {\bf 349} (2004) 1039--1051,
  [\href{http://xxx.lanl.gov/abs/astro-ph/0311231}{{\tt astro-ph/0311231}}].

\bibitem{Navarro:2008kc}
J.~F. Navarro, A.~Ludlow, {\em et.~al.}, {\it {The Diversity and Similarity of
  Cold Dark Matter Halos}},  \href{http://xxx.lanl.gov/abs/0810.1522}{{\tt
  arXiv:0810.1522}}.

\bibitem{Persic:1995ru}
M.~Persic, P.~Salucci, and F.~Stel, {\it {The Universal rotation curve of
  spiral galaxies: 1. The Dark matter connection}},  {\em
  Mon.Not.Roy.Astron.Soc.} {\bf 281} (1996) 27,
  [\href{http://xxx.lanl.gov/abs/astro-ph/9506004}{{\tt astro-ph/9506004}}].

\bibitem{Gentile:2004tb}
G.~Gentile, P.~Salucci, U.~Klein, D.~Vergani, and P.~Kalberla, {\it {The Cored
  distribution of dark matter in spiral galaxies}},  {\em
  Mon.Not.Roy.Astron.Soc.} {\bf 351} (2004) 903,
  [\href{http://xxx.lanl.gov/abs/astro-ph/0403154}{{\tt astro-ph/0403154}}].

\bibitem{gondolosilk}
P.~{Gondolo} and J.~{Silk}, {\it {Dark Matter Annihilation at the Galactic
  Center}},  {\em \prl} {\bf 83} (1999) 1719--1722,
  [\href{http://xxx.lanl.gov/abs/astro-ph/9906391}{{\tt astro-ph/9906391}}].

\bibitem{Gnedin:2004cx}
O.~Y. Gnedin, A.~V. Kravtsov, A.~A. Klypin, and D.~Nagai, {\it {Response of
  dark matter halos to condensation of baryons: Cosmological simulations and
  improved adiabatic contraction model}},  {\em Astrophys.J.} {\bf 616} (2004)
  16--26, [\href{http://xxx.lanl.gov/abs/astro-ph/0406247}{{\tt
  astro-ph/0406247}}].

\bibitem{Gnedin:2011uj}
O.~Y. Gnedin, D.~Ceverino, {\em et.~al.}, {\it {Halo Contraction Effect in
  Hydrodynamic Simulations of Galaxy Formation}},
  \href{http://xxx.lanl.gov/abs/1108.5736}{{\tt arXiv:1108.5736}}.

\bibitem{Prada:2004pi}
F.~Prada, A.~Klypin, J.~Flix~Molina, M.~Martinez, and E.~Simonneau, {\it {Dark
  Matter Annihilation in the Milky Way Galaxy: Effects of Baryonic
  Compression}},  {\em Phys.Rev.Lett.} {\bf 93} (2004) 241301,
  [\href{http://xxx.lanl.gov/abs/astro-ph/0401512}{{\tt astro-ph/0401512}}].

\bibitem{Gustafsson:2006gr}
M.~Gustafsson, M.~Fairbairn, and J.~Sommer-Larsen, {\it {Baryonic Pinching of
  Galactic Dark Matter Haloes}},  {\em Phys.Rev.} {\bf D74} (2006) 123522,
  [\href{http://xxx.lanl.gov/abs/astro-ph/0608634}{{\tt astro-ph/0608634}}].

\bibitem{Lacroix:2013qka}
T.~Lacroix, C.~Boehm, and J.~Silk, {\it {Probing a dark matter density spike at
  the Galactic Center}},  \href{http://xxx.lanl.gov/abs/1311.0139}{{\tt
  arXiv:1311.0139}}.

\bibitem{Mashchenko:2007jp}
S.~Mashchenko, J.~Wadsley, and H.~Couchman, {\it {Stellar Feedback in Dwarf
  Galaxy Formation}},  {\em Science} {\bf 319} (2008) 174,
  [\href{http://xxx.lanl.gov/abs/0711.4803}{{\tt arXiv:0711.4803}}].

\bibitem{Maccio:2011eh}
A.~V. Maccio', G.~Stinson, {\em et.~al.}, {\it {Halo expansion in cosmological
  hydro simulations: towards a baryonic solution of the cusp/core problem in
  massive spirals}},  {\em Astrophys. J. Lett.} {\bf 744} (2012) L9,
  [\href{http://xxx.lanl.gov/abs/1111.5620}{{\tt arXiv:1111.5620}}].

\bibitem{Pontzen:2011ty}
A.~Pontzen and F.~Governato, {\it {How supernova feedback turns dark matter
  cusps into cores}},  {\em Mon. Not. Roy. Astron. Soc.} {\bf 421} (2012) 3464,
  [\href{http://xxx.lanl.gov/abs/1106.0499}{{\tt arXiv:1106.0499}}].

\bibitem{Governato:2012fa}
F.~Governato, A.~Zolotov, {\em et.~al.}, {\it {Cuspy No More: How Outflows
  Affect the Central Dark Matter and Baryon Distribution in Lambda CDM
  Galaxies}},  {\em Mon.Not.Roy.Astron.Soc.} {\bf 422} (2012) 1231--1240,
  [\href{http://xxx.lanl.gov/abs/1202.0554}{{\tt arXiv:1202.0554}}].

\bibitem{Rocha:2012jg}
M.~Rocha, A.~H. Peter, {\em et.~al.}, {\it {Cosmological Simulations with
  Self-Interacting Dark Matter I: Constant Density Cores and Substructure}},
  {\em Mon.Not.Roy.Astron.Soc.} {\bf 430} (2013) 81--104,
  [\href{http://xxx.lanl.gov/abs/1208.3025}{{\tt arXiv:1208.3025}}].

\bibitem{Kaplinghat:2013xca}
M.~Kaplinghat, R.~E. Keeley, T.~Linden, and H.-B. Yu, {\it {Tying Dark Matter
  to Baryons with Self-interactions}},
  \href{http://xxx.lanl.gov/abs/1311.6524}{{\tt arXiv:1311.6524}}.

\bibitem{Ripken11}
J.~{Ripken}, J.~{Conrad}, and P.~{Scott}, {\it {Implications for constrained
  supersymmetry of combined H.E.S.S. observations of dwarf galaxies, the
  Galactic halo and the Galactic centre}},  {\em \jcap} {\bf 11} (2011) 004,
  [\href{http://xxx.lanl.gov/abs/1012.3939}{{\tt arXiv:1012.3939}}].

\bibitem{Acharya:2013sxa}
B.~Acharya, M.~Actis, {\em et.~al.}, {\it {Introducing the CTA concept}},  {\em
  Astropart.Phys.} {\bf 43} (2013) 3--18.

\bibitem{Bergstrom11}
L.~{Bergstr{\"o}m}, T.~{Bringmann}, and J.~{Edsj{\"o}}, {\it {Complementarity
  of direct dark matter detection and indirect detection through gamma rays}},
  {\em \prd} {\bf 83} (2011) 045024,
  [\href{http://xxx.lanl.gov/abs/1011.4514}{{\tt arXiv:1011.4514}}].

\bibitem{Doro:2012xx}
CTA collaboration: M.~Doro {\em et.~al.}, {\it {Dark Matter and Fundamental
  Physics with the Cherenkov Telescope Array}},  {\em Astropart.Phys.} {\bf 43}
  (2013) 189--214, [\href{http://xxx.lanl.gov/abs/1208.5356}{{\tt
  arXiv:1208.5356}}].

\bibitem{Wood:2013taa}
M.~Wood, J.~Buckley, {\em et.~al.}, {\it {Prospects for Indirect Detection of
  Dark Matter with CTA}},  \href{http://xxx.lanl.gov/abs/1305.0302}{{\tt
  arXiv:1305.0302}}.

\bibitem{Cline13b}
J.~M. {Cline}, K.~{Kainulainen}, P.~{Scott}, and C.~{Weniger}, {\it {Update on
  scalar singlet dark matter}},  {\em \prd} {\bf 88} (2013) 055025,
  [\href{http://xxx.lanl.gov/abs/1306.4710}{{\tt arXiv:1306.4710}}].

\bibitem{Aguilar:2013qda}
AMS Collaboration: M.~Aguilar {\em et.~al.}, {\it {First Result from the Alpha
  Magnetic Spectrometer on the International Space Station: Precision
  Measurement of the Positron Fraction in Primary Cosmic Rays of 0.5--350
  GeV}},  {\em Phys.Rev.Lett.} {\bf 110} (2013) 141102.

\bibitem{Adriani:2008zr}
PAMELA Collaboration: O.~Adriani {\em et.~al.}, {\it {An anomalous positron
  abundance in cosmic rays with energies 1.5-100 GeV}},  {\em Nature} {\bf 458}
  (2009) 607--609, [\href{http://xxx.lanl.gov/abs/0810.4995}{{\tt
  arXiv:0810.4995}}].

\bibitem{FermiLAT:2011ab}
Fermi LAT Collaboration: M.~Ackermann {\em et.~al.}, {\it {Measurement of
  separate cosmic-ray electron and positron spectra with the Fermi Large Area
  Telescope}},  {\em Phys.Rev.Lett.} {\bf 108} (2012) 011103,
  [\href{http://xxx.lanl.gov/abs/1109.0521}{{\tt arXiv:1109.0521}}].

\bibitem{Chang:2008aa}
J.~Chang, J.~Adams, {\em et.~al.}, {\it {An excess of cosmic ray electrons at
  energies of 300-800 GeV}},  {\em Nature} {\bf 456} (2008) 362--365.

\bibitem{Abdo:2009zk}
Fermi LAT Collaboration: A.~A. Abdo {\em et.~al.}, {\it {Measurement of the
  Cosmic Ray e+ plus e- spectrum from 20 GeV to 1 TeV with the Fermi Large Area
  Telescope}},  {\em Phys.Rev.Lett.} {\bf 102} (2009) 181101,
  [\href{http://xxx.lanl.gov/abs/0905.0025}{{\tt arXiv:0905.0025}}].

\bibitem{Aharonian:2008aa}
H.E.S.S. Collaboration: F.~Aharonian {\em et.~al.}, {\it {The energy spectrum
  of cosmic-ray electrons at TeV energies}},  {\em Phys.Rev.Lett.} {\bf 101}
  (2008) 261104, [\href{http://xxx.lanl.gov/abs/0811.3894}{{\tt
  arXiv:0811.3894}}].

\bibitem{Aharonian:2009ah}
H.E.S.S. Collaboration: F.~Aharonian {\em et.~al.}, {\it {Probing the ATIC peak
  in the cosmic-ray electron spectrum with H.E.S.S}},  {\em Astron.Astrophys.}
  {\bf 508} (2009) 561, [\href{http://xxx.lanl.gov/abs/0905.0105}{{\tt
  arXiv:0905.0105}}].

\bibitem{Cirelli:2012ut}
M.~Cirelli, E.~Moulin, P.~Panci, P.~D. Serpico, and A.~Viana, {\it {Gamma ray
  constraints on Decaying Dark Matter}},  {\em Phys.Rev.} {\bf D86} (2012)
  083506, [\href{http://xxx.lanl.gov/abs/1205.5283}{{\tt arXiv:1205.5283}}].

\bibitem{Abazajian:2011ak}
K.~N. Abazajian and J.~P. Harding, {\it {Constraints on WIMP and
  Sommerfeld-Enhanced Dark Matter Annihilation from HESS Observations of the
  Galactic Center}},  {\em JCAP} {\bf 1201} (2012) 041,
  [\href{http://xxx.lanl.gov/abs/1110.6151}{{\tt arXiv:1110.6151}}].

\bibitem{Abdo:2010dk}
Fermi-LAT Collaboration: A.~Abdo {\em et.~al.}, {\it {Constraints on
  Cosmological Dark Matter Annihilation from the Fermi-LAT Isotropic Diffuse
  Gamma-Ray Measurement}},  {\em JCAP} {\bf 1004} (2010) 014,
  [\href{http://xxx.lanl.gov/abs/1002.4415}{{\tt arXiv:1002.4415}}].

\bibitem{Ackermann:2012rg}
LAT collaboration: M.~Ackermann {\em et.~al.}, {\it {Constraints on the
  Galactic Halo Dark Matter from Fermi-LAT Diffuse Measurements}},  {\em
  Astrophys.J.} {\bf 761} (2012) 91,
  [\href{http://xxx.lanl.gov/abs/1205.6474}{{\tt arXiv:1205.6474}}].

\bibitem{Bringmann:2011ye}
T.~Bringmann, F.~Calore, G.~Vertongen, and C.~Weniger, {\it {On the Relevance
  of Sharp Gamma-Ray Features for Indirect Dark Matter Searches}},  {\em
  Phys.Rev.} {\bf D84} (2011) 103525,
  [\href{http://xxx.lanl.gov/abs/1106.1874}{{\tt arXiv:1106.1874}}].

\bibitem{Aleksic:2013xea}
J.~Aleksi{\'c}, S.~Ansoldi, {\em et.~al.}, {\it {Optimized dark matter searches
  in deep observations of Segue 1 with MAGIC}},
  \href{http://xxx.lanl.gov/abs/1312.1535}{{\tt arXiv:1312.1535}}.

\bibitem{Gondolo:2004sc}
P.~Gondolo, J.~Edsjo, {\em et.~al.}, {\it {DarkSUSY: Computing supersymmetric
  dark matter properties numerically}},  {\em JCAP} {\bf 0407} (2004) 008,
  [\href{http://xxx.lanl.gov/abs/astro-ph/0406204}{{\tt astro-ph/0406204}}].

\bibitem{Salucci:2010qr}
P.~Salucci, F.~Nesti, G.~Gentile, and C.~Martins, {\it {The dark matter density
  at the Sun's location}},  {\em Astron.Astrophys.} {\bf 523} (2010) A83,
  [\href{http://xxx.lanl.gov/abs/1003.3101}{{\tt arXiv:1003.3101}}].

\bibitem{Bernlohr:2012we}
K.~Bernl{\"o}hr, A.~Barnacka, {\em et.~al.}, {\it {Monte Carlo design studies
  for the Cherenkov Telescope Array}},  {\em Astropart.Phys.} {\bf 43} (2013)
  171--188, [\href{http://xxx.lanl.gov/abs/1210.3503}{{\tt arXiv:1210.3503}}].

\bibitem{Mack:2008wu}
G.~D. Mack, T.~D. Jacques, J.~F. Beacom, N.~F. Bell, and H.~Yuksel, {\it
  {Conservative Constraints on Dark Matter Annihilation into Gamma Rays}},
  {\em Phys.Rev.} {\bf D78} (2008) 063542,
  [\href{http://xxx.lanl.gov/abs/0803.0157}{{\tt arXiv:0803.0157}}].

\bibitem{Skellam}
J.~G. {Skellam}, {\it The frequency distribution of the difference between two
  poisson variates belonging to different populations},  {\em J.\ Royal Stat.\
  Soc.} {\bf 109} (1946) 296.

\bibitem{Abramowski:2011hc}
H.E.S.S.Collaboration: A.~Abramowski {\em et.~al.}, {\it {Search for a Dark
  Matter annihilation signal from the Galactic Center halo with H.E.S.S}},
  {\em Phys.Rev.Lett.} {\bf 106} (2011) 161301,
  [\href{http://xxx.lanl.gov/abs/1103.3266}{{\tt arXiv:1103.3266}}].

\bibitem{Abdo:2010nz}
Fermi-LAT collaboration: A.~Abdo {\em et.~al.}, {\it {The Spectrum of the
  Isotropic Diffuse Gamma-Ray Emission Derived From First-Year Fermi Large Area
  Telescope Data}},  {\em Phys.Rev.Lett.} {\bf 104} (2010) 101101,
  [\href{http://xxx.lanl.gov/abs/1002.3603}{{\tt arXiv:1002.3603}}].

\end{thebibliography}\endgroup

\end{document}